\documentclass[english]{appolb}
\usepackage[T1]{fontenc}
\usepackage[latin9]{inputenc}
\usepackage[paperwidth=23cm,paperheight=30cm]{geometry}
\usepackage{color}
\usepackage{babel}
\usepackage{array}
\usepackage{float}
\usepackage{textcomp}
\usepackage{multirow}
\usepackage{amsthm}
\usepackage{amsmath}
\usepackage{graphicx}
\usepackage{esint}
\PassOptionsToPackage{normalem}{ulem}
\usepackage{ulem}
\usepackage[unicode=true,
 bookmarks=true,bookmarksnumbered=true,bookmarksopen=true,bookmarksopenlevel=1,
 breaklinks=false,pdfborder={0 0 1},backref=false,colorlinks=true]
 {hyperref}
\hypersetup{pdftitle={Comparison of Cut-Based and Matrix Element Method Results for Beyond Standard Model Quarks},
 pdfauthor={E. Akyazi, G. Unel, M. Yilmaz}}

\makeatletter

\providecommand{\tabularnewline}{\\}

\numberwithin{equation}{section}
\numberwithin{figure}{section}
\newcommand{\lyxaddress}[1]{
\par {\raggedright #1
\vspace{1.4em}
\noindent\par}
}

\makeatother

\begin{document}

\title{Comparison of Cut-Based and Matrix Element Method Results for Beyond
Standard Model Quarks}

\author{E. Akyazi$^{1}$%
\thanks{akyazi@cern.ch%
}, G. Unel$^{2}$%
\thanks{Gokhan.Unel@cern.ch%
}, M. Yilmaz$^{3}$ %
\thanks{metiny@gazi.edu.tr%
}}

\maketitle

\lyxaddress{$^{1,3}$ Gazi University, Department of Physics, Ankara, \textcolor{black}{Turkey.
}\\
\textcolor{black}{$^{2}$ University of California at Irvine, Physics
Department, USA. }}
\begin{abstract}
In this work, two different methods for extracting the mass of a new
quark from the (pseudo) data are compared: the classical cut-based
method and the matrix element method. As a concrete example, a fourth
family up type quark is searched in p-p collisions of 7 TeV center
of mass energy. We have shown that even with a very small number of
events, Matrix Element Method gives better estimations for the mass
value and its error. Especially, for event samples in which Signal
to Background ratio is greater than 0.2 Matrix Element Method reduces
the statistical error approximately ten times.
\end{abstract}

\section{Introduction}

In searching for new phenomena at the particle physics experiments,
it is very important to extract the values of the unknown parameters
with maximal statistical significance from small data samples. At
this point, Matrix Element Method (MEM) provides a very powerful tool
which gave the most precise value for top quark mass at Tevatron experiments
DØ and CDF {[}1,2,3,4{]}. After the method became more popular, it
has also been applied to other analysis such as electroweak single
top quark production {[}5{]}, estimation of the longitudinal W boson
helicity fraction in top quark decays {[}6{]} and searches for the
Higgs boson {[}7{]}. It can be applied to any mass analysis which
includes exclusive decay channels at hadron colliders for BSM researches.
In this paper, a brief description of this method is followed by a
comparison of the results of a heavy quark search analysis using a
traditional cut-based method, to those from the Matrix Element Method.

\subsection{Matrix Element Method: }

The name Matrix Element Method comes from the fact that probability
function which is used in this method is driven by the physical matrix
element. Matrix Element Method uses both theoretical and experimental
information to extract the values of any unknown parameters from the
experimental data. Therefore, the essential point of the MEM is that,
it maximally uses the information contained in the physics of the
problem, without trying to extract it from the distributions as in
the case of cut and count method. In this technique, each experimentally
measured quantity is associated to a Bayesian probability function
P($x$|$\alpha$) which gives the probability to observe this event
in a certain theoretical frame $\alpha$. The probability weight which
is based on square matrix element {[}8,9{]} can be written in the
following form {[}10,11{]}:

\begin{equation}
P(x|\alpha)=\frac{1}{\sigma}\int d\phi(y)|M|^{2}dw_{1}dw_{2}f_{1}(w_{1})f_{2}(w_{2})W(x,y)\label{eq:olasilik}
\end{equation}

~

\noindent where $x$ is a set of detector-level kinematic quantities,
\textit{y} is the parton-level 4-vectors, $\sigma$ is the parton
level cross-section (1/$\sigma$ factor ensures the normalization
of probability), $M$ is the matrix element describing the production
and decay process, $f_{1}(w_{1})$ and $f_{2}(w_{2})$ are parton
distribution functions, $d\phi(y)$ is phase-space element and $W(x,y)$
is the transfer function or resolution function which describes the
probability density to reconstruct an assumed partonic final state
\textit{y} as a measurement $x$ in the detector.

~

The probability is derived by integrating over all possible parton
states, and each configuration is weighted according to its probability
to produce the observed measurement. The weights are then combined
together into a likelihood to determine the most probable value of
the parameter of interest (top quark mass, W helicity, etc). 

~

The likelihood function for N measured events can be written as: 

\begin{equation}
L(\alpha)=e^{-N\int\bar{P}(x,\alpha)dx}\prod_{i=1}^{N}\bar{P}(x_{i},\alpha)\label{eq:likel}
\end{equation}

~

\noindent where $\alpha$ is any parameter that we want to estimate
and $\bar{P}(x_{i},\alpha)$ is measured probability density. The
derivation of likelihood can be found in {[}12{]}. The best value
of $\alpha$ is obtained through maximization of the likelihood or
more practically by minimizing -\textit{ln L($\alpha$)} with respect
to $\alpha$.

\subsubsection{Transfer Functions: }

The determination of transfer functions (TF) is the most important
part of Matrix Element Method. As mentioned before, transfer functions
map parton level quantities to detector level measured quantities
or vice versa. The energy resolution of the leptons and the jets is
parametrized with transfer functions W($\Delta E=E_{parton}-E_{jet}$)
and they give the probability for a measurement E$_{jet}$ in the
detector, if the true object energy is E$_{parton}$. TFs can be decomposed
into a product of functions for each external or internal particle,
and each part can be handled separately. Although there are different
type of TFs that can be found in various analysis, the most used one
for jets is Cannelli's double gaussian formulation {[}13{]}: one gaussian
is for the symmetric peak while the other one accommodates the asymmetric
tails of the $\bigtriangleup E$ distribution. In this formulation
jet transfer function is expressed to be a function only of the relative
energy difference between the parton and the jet :

\begin{gather}
W(\Delta E)=\frac{1}{\sqrt{2\pi}(a_{2}+a_{3}a_{5})}(exp(-\frac{(\Delta E-a_{1})^{2}}{2a_{2}^{2}})+a_{3}\, exp(-\frac{(\Delta E-a_{4})^{2}}{2a_{5}^{2}}))\label{eq:transferfonk}
\end{gather}

~

\noindent where the energy dependence of these $a{}_{i}$ parameters
can be written in following form {[}14{]}:
\begin{eqnarray}
a_{i} & = & a_{i,0}+a_{i,1}\sqrt{E}+a_{i,2}E\label{eq:tfparams}
\end{eqnarray}

These parameters can be determined by minimizing a likelihood formed
by measuring parton energy and matched jet energy in a Monte Carlo
sample under consideration and they must be determined in different
pseudorapidity regions of the calorimeter to account for resolution
differences in the detector. 

Theoretically lepton energies and angles can be parametrized as a
gaussian but in practice they are assumed to be almost well-measured
by a detector apparatus, so the TFs for lepton energies and all the
particle angles can be parametrized by delta functions. This parametrization
is also less time-consuming for computation of the weights because
of the dimensional reduction it introduces.

\section{Analysis:}

In this work, comparison of Matrix Element Method and cut-based method
for mass reconstruction analysis of fourth family up type quark {[}15{]},
\textit{u$_{4}$,} at 7 TeV center of mass energy using event samples
which include different Signal to Background (S/B) ratios has been
presented. For simplicity, neither detector resolution effects nor
systematic effects haven't been considered in this study.

This analysis is based on Monte Carlo events generated with MadGraph/MadEvent
{[}16{]} and processed through Pythia {[}17{]} for the parton-shower
and hadronization. Finally, detector response is simulated by PGS
{[}18{]}. In this study, the mixing between fourth generation and
the first SM family is assumed to be 100 percent. Therefore, the decay
channel $u{}_{4}$ \textrightarrow{}\textit{ W}+ \textit{d} becomes
the dominant one. As signal, the pair production of up type fourth
family quark, $u{}_{4}$, at a proton-proton collider at a center
of mass energy of 7 TeV is considered. The full process for signal
events can be written as:

\textit{
\begin{equation}
\centering pp\rightarrow u_{4}\bar{u}_{4}\rightarrow W^{-}W^{+}jj\label{eq:surec}
\end{equation}
}where \textit{$j$} is a jet originating from a \textit{d} quark
or $\bar{d}$ quark and one \textit{W} decays leptonically whereas
the other decays hadronically. For simplicity, electronic decay mode
of the \textit{W} is considered. Therefore, the signal is searched
in the\textit{ 4j+1e+MET} final state. As the dominant background
sample, \textit{t}$\bar{t}$ events in which the top quark pairs decay
semi-leptonically has been taken under consideration. These background
events are also produced with MG-ME/Pythia-PGS chain with CTEQ6L1
{[}19{]} as the PDF set.

The Monte Carlo events have been produced for three different mass
values of \textit{u$_{4}$} quark: 400, 500 and 600 GeV. These events
were required to contain the right number of jets and leptons in the
final state (i.e. 4 jets and 1 electron for this study).

\subsection{Cut-Based Analysis:}

In the cut-based analysis, leptonically decaying \textit{W} bosons
were reconstructed from the 4-momentum of the lepton and the missing
transverse momentum. Assuming a massless neutrino and on-shell W mass,
the $z$ component of the neutrino, and its energy are obtained by
solving these two equations with two unknowns. If the equations can
be solved, the solution providing the smallest $\left|P_{z}\right|$
is selected. The rational behind this selection is to use the smallest
estimated value, thus to reduce the error margin. If the equation
set cannot be solved ($\Delta<0$) then, the neutrino four momentum
is formed using the collinearity approximation, i.e. by assuming the
same $\eta$ for the neutral and charged leptons and again a massless
neutrino. Hadronically decaying \textit{W} bosons were reconstructed
using the 4-momentum of two soft jets in each event. The two relevant
jets are selected by considering the pairing of all jets, and by selecting
the pair which would minimize a $\chi^{2}$ defined as:

~

\begin{equation}
\chi^{2}\equiv\frac{(M_{jj}-M_{_{W}})^{2}}{\sigma_{W}^{2}}+\frac{(M_{jjj}-M_{j\nu l})^{2}}{\sigma_{Q}^{2}}\label{eq:kikare}
\end{equation}

where $M_{jj}$ is the reconstructed invariant mass from two jets,
$M_{jjj}$ is the reconstructed invariant mass from three jets, $M_{j\nu l}$
is reconstructed invariant mass from lepton, MET and jet, $\sigma_{W}$
is decay width of W, $\sigma_{Q}$ is decay width of new heavy quark.
The W-jet association ambiguity is resolved by selecting the combination
which yields in the smallest difference between the masses of the
two reconstructed \textit{u$_{4}$} quarks in the same event. The
\textit{u$_{4}$} invariant mass is obtained by taking the average
of the hadronically and leptonically decaying \textit{u$_{4}$} quarks.
In the generation step, standard kinematic selection criteria are
applied as follows:
\begin{align}
P_{T,e} & >10GeV,\nonumber \\
P_{T,j} & >20GeV,\nonumber \\
|\eta_{e}| & <2.5\,,\nonumber \\
\triangle R(e,j) & >0.4,\nonumber \\
\triangle R(j,j) & >0.4,\nonumber \\
|\eta_{j}| & <5.\label{ec:TTdec}
\end{align}

where \textit{P$_{T,e}$ (P$_{T,j}$} \textit{)} is transverse momentum
of electrons (jets), $|\eta_{e}$| ($|\eta_{j}|$), is the rapidity
for electrons (jets) and, $\Delta R\left(e,j\right)$ is the angular
distance between electrons and jets and $\Delta R\left(j,j\right)$
is angular distance between jets, with $\bigtriangleup R\equiv\sqrt{\bigtriangleup\eta^{2}+\bigtriangleup\phi^{2}}$
.

\begin{figure}[H]
\centering\includegraphics[scale=0.6]{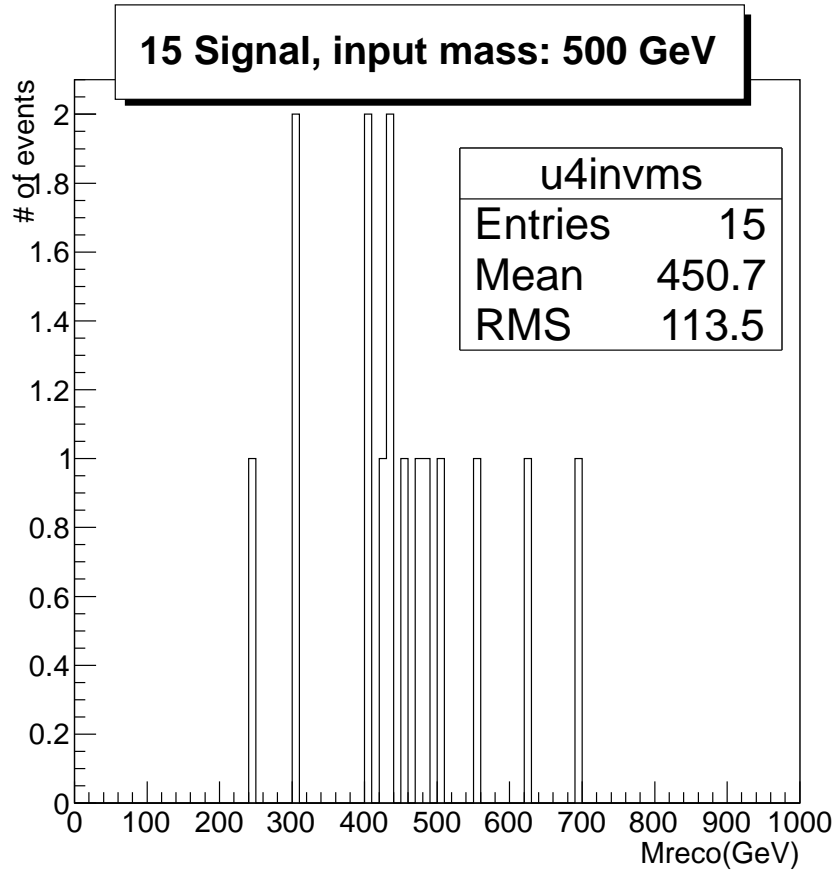}\caption{Invariant mass histogram of \textit{u$_{4}$} with the cut-based method
for an input test mass of 500 GeV. The result is extracted from a
pure signal sample which contain only 15 events.\label{fig:Invariant-mass-histogram}}
\end{figure}

In the first reconstruction step, \textit{a u$_{4}$} quark of 500
GeV was used and \textit{u$_{4}$} invariant mass was extracted from
a sample containing only 15 signal events. The reconstructed mass
histogram for this case shown in Fig. \ref{fig:Invariant-mass-histogram}.

The same procedure has been applied to other samples containing different
numbers of signal and background events. In short, the S/B ratio was
scanned from a purely signal sample down to a purely background sample
keeping the total number of events same, namely, 15. The cases which
were scanned are: 13 signal (S) + 2 background (B), 11 S + 4 B, 9
S+ 6 B, 7 S + 8 B, 5 S +10 B, 3 S + 12 B, 15 B. Invariant mass histograms
obtained for these cases are shown in Fig. \ref{fig:500gevcbresults}.
\begin{quotation}
\begin{figure}[H]
\centering\includegraphics[scale=0.27]{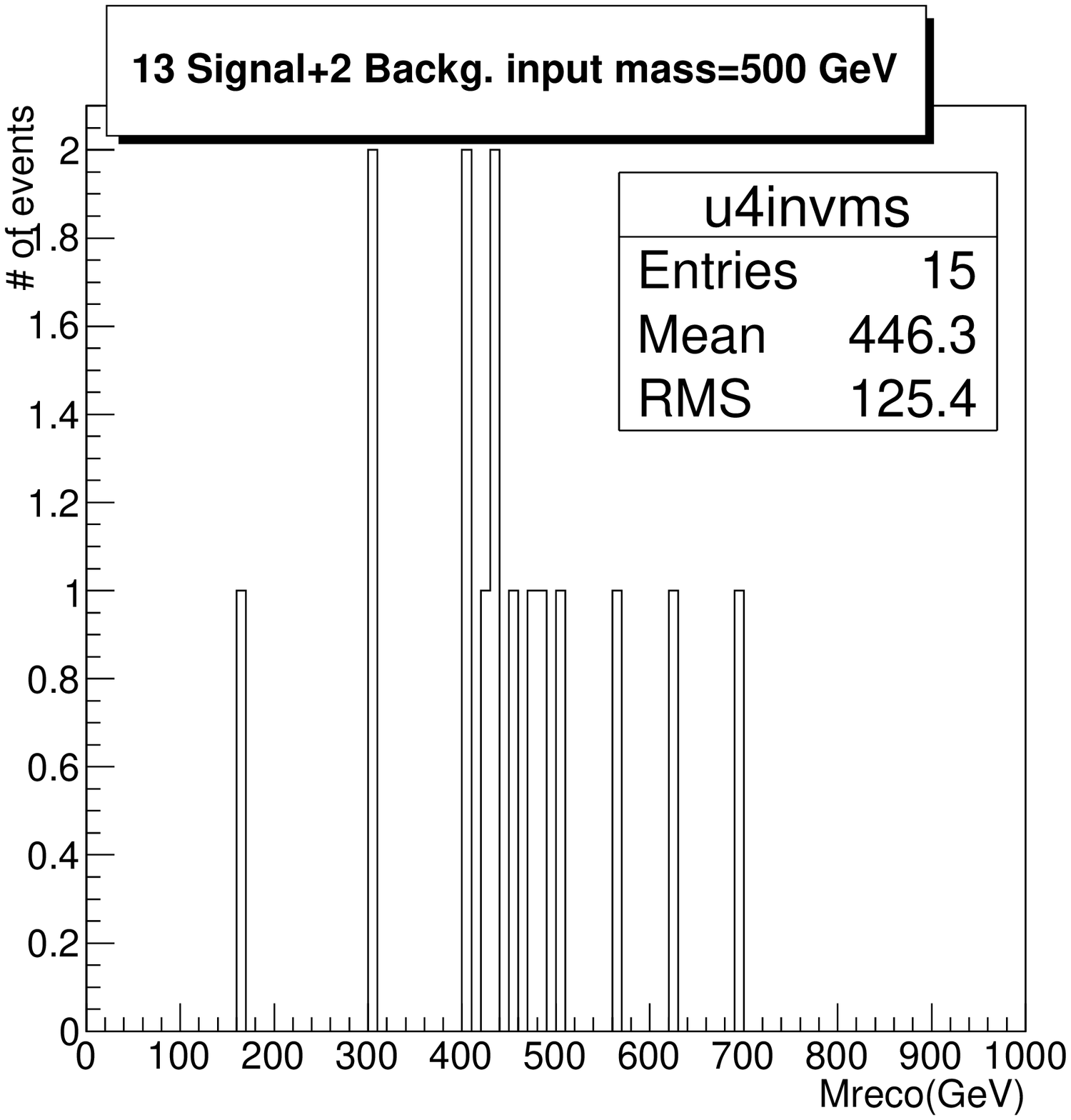}\includegraphics[scale=0.27]{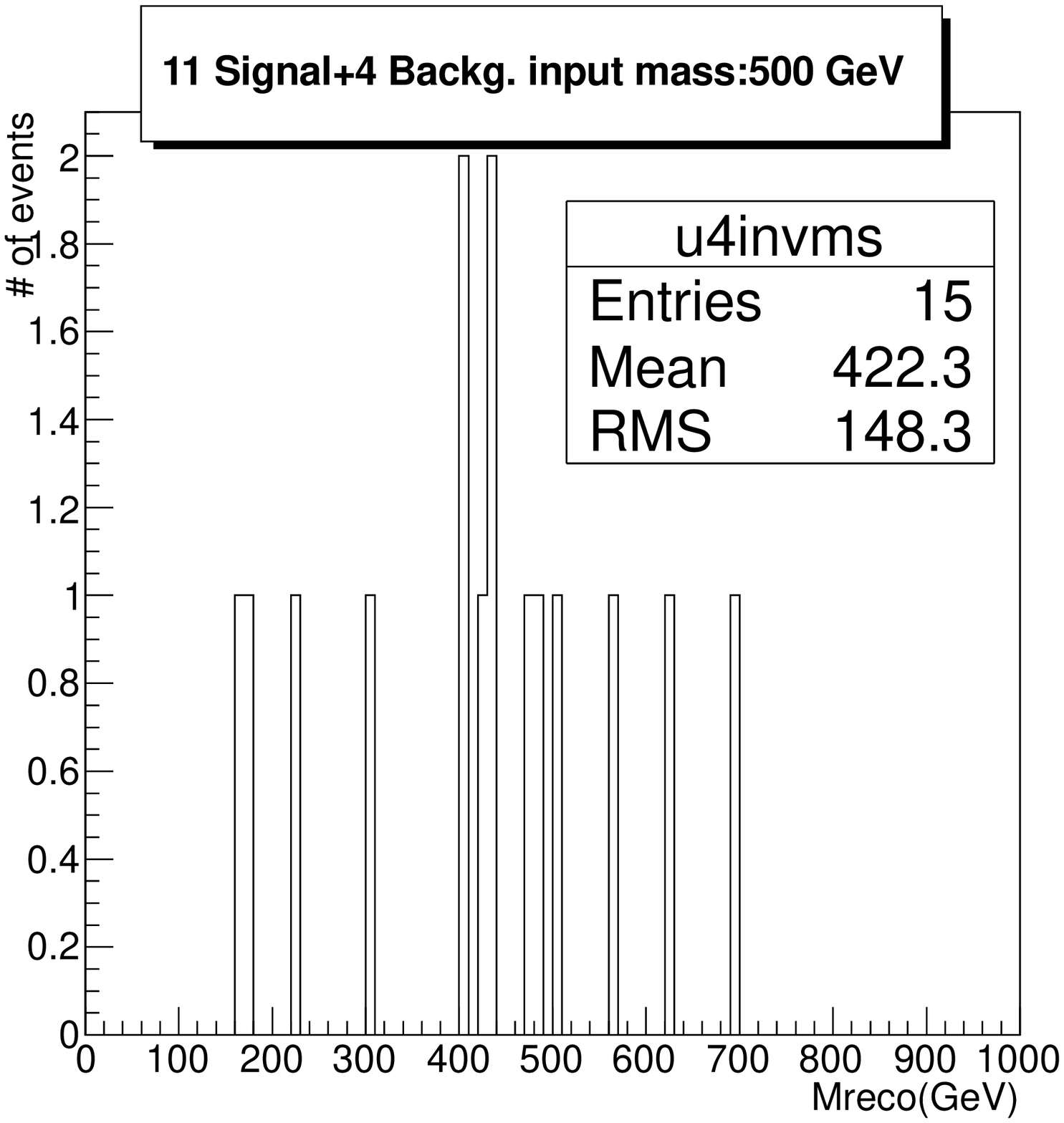}

~

\includegraphics[scale=0.27]{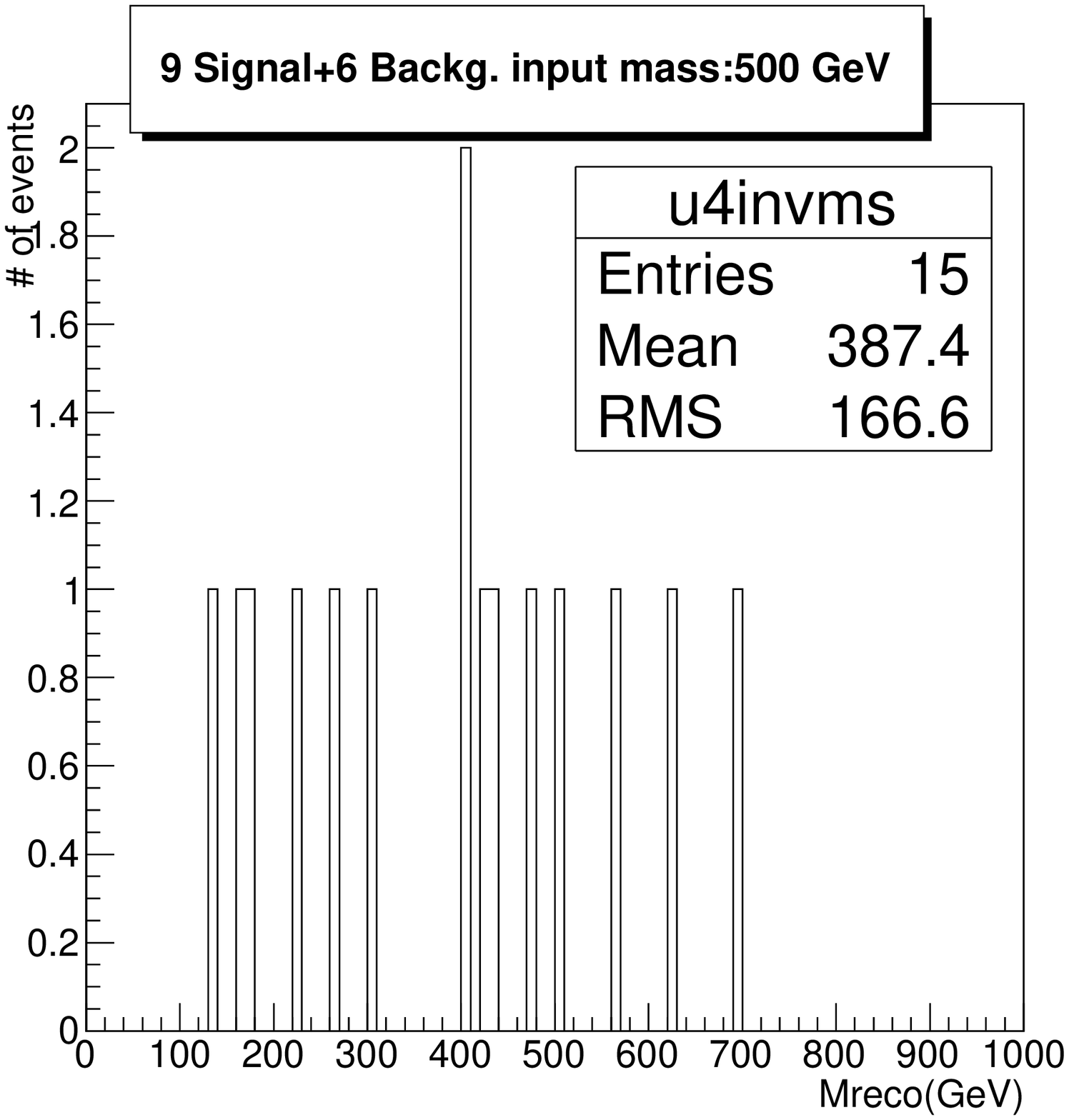}\includegraphics[scale=0.27]{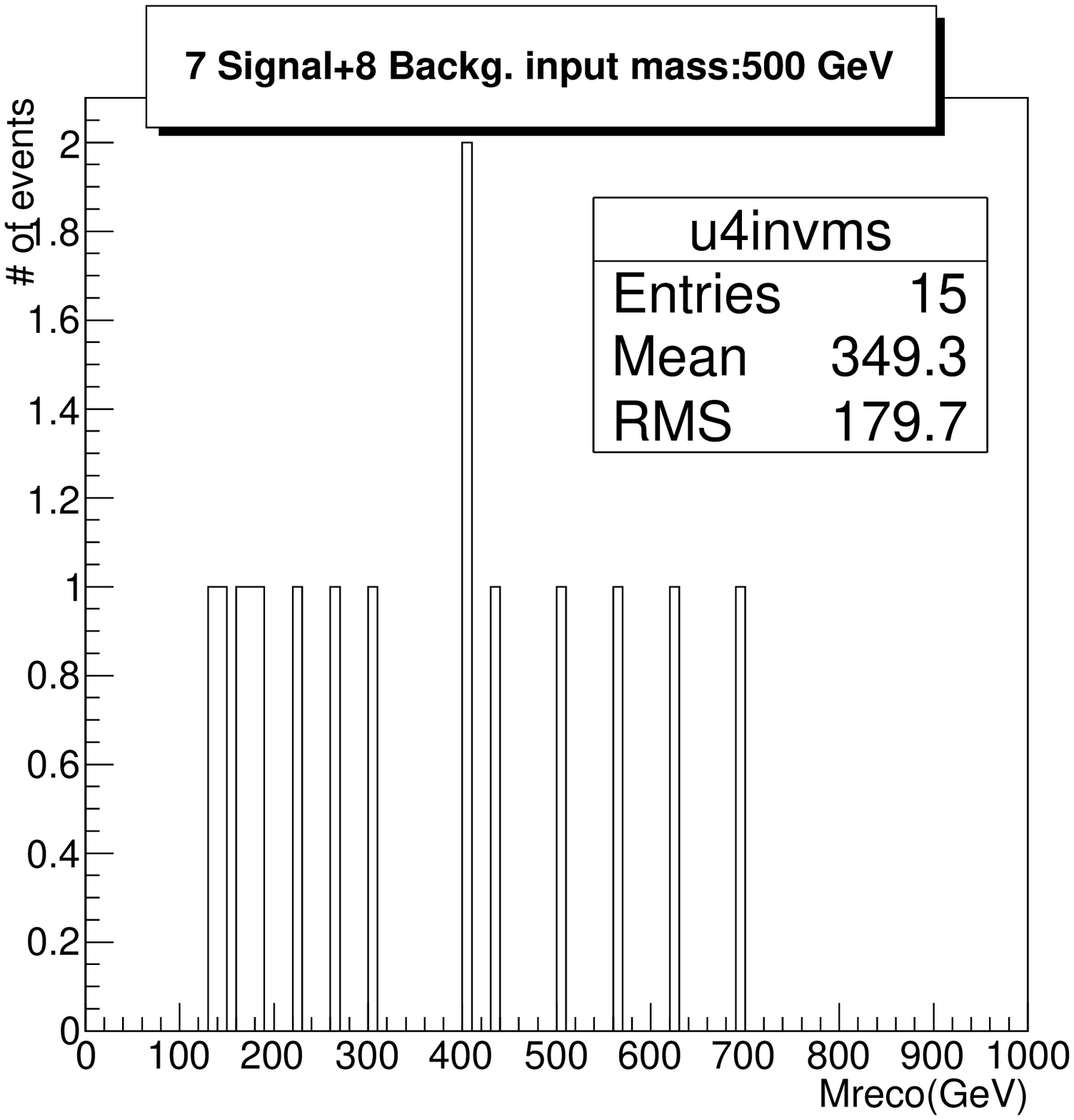}

~

\includegraphics[scale=0.27]{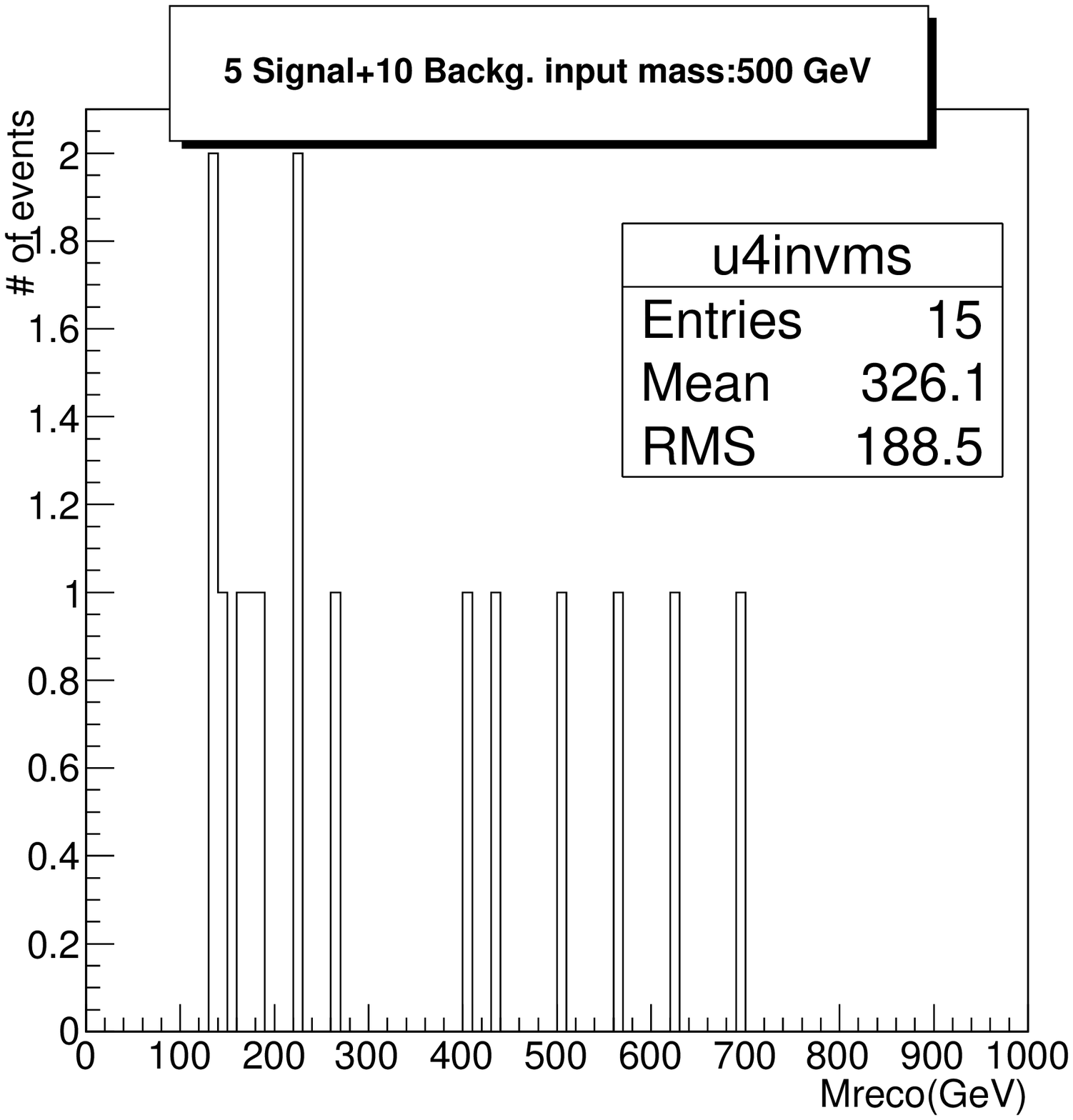}\includegraphics[scale=0.27]{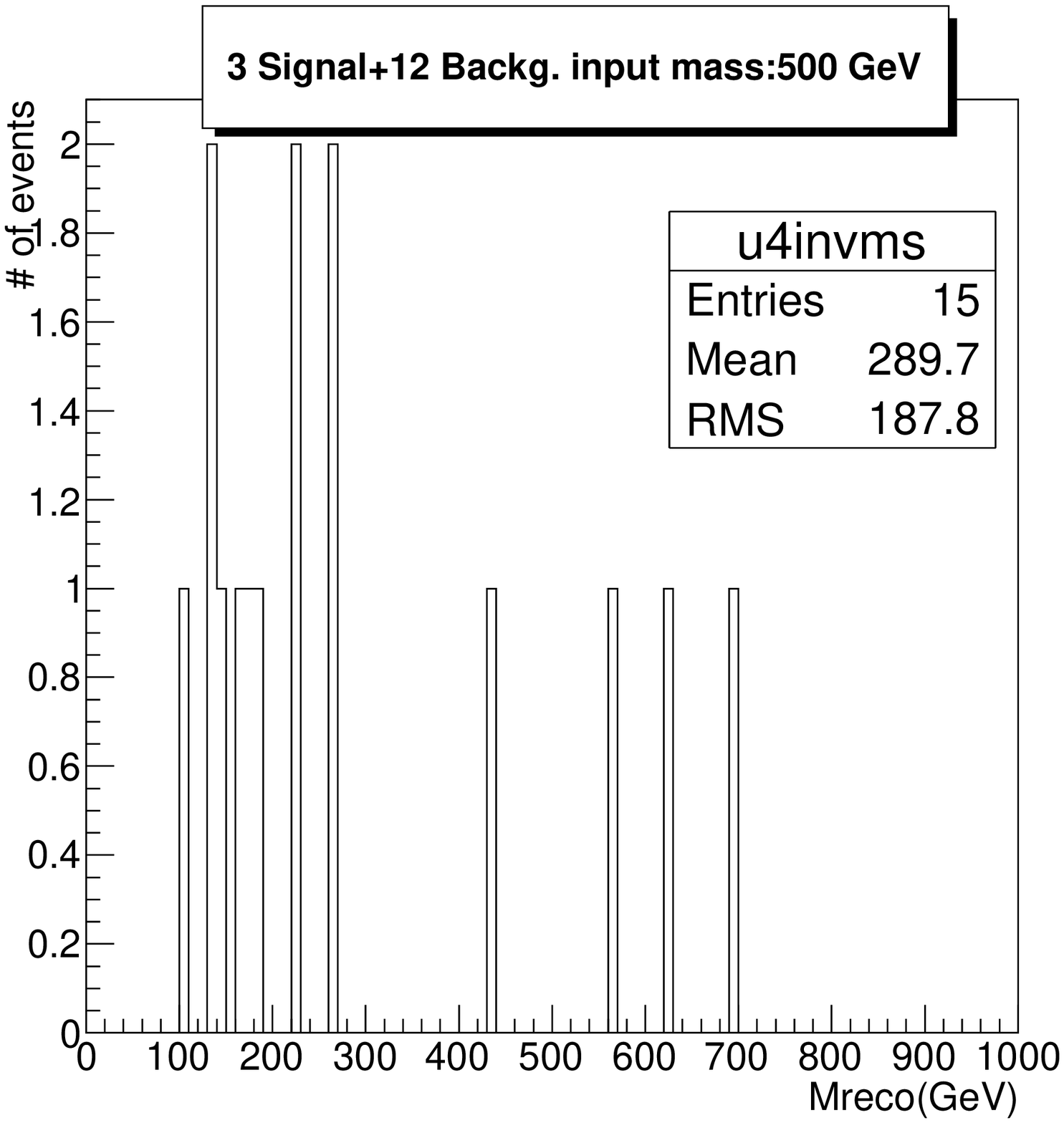}

~\includegraphics[scale=0.28]{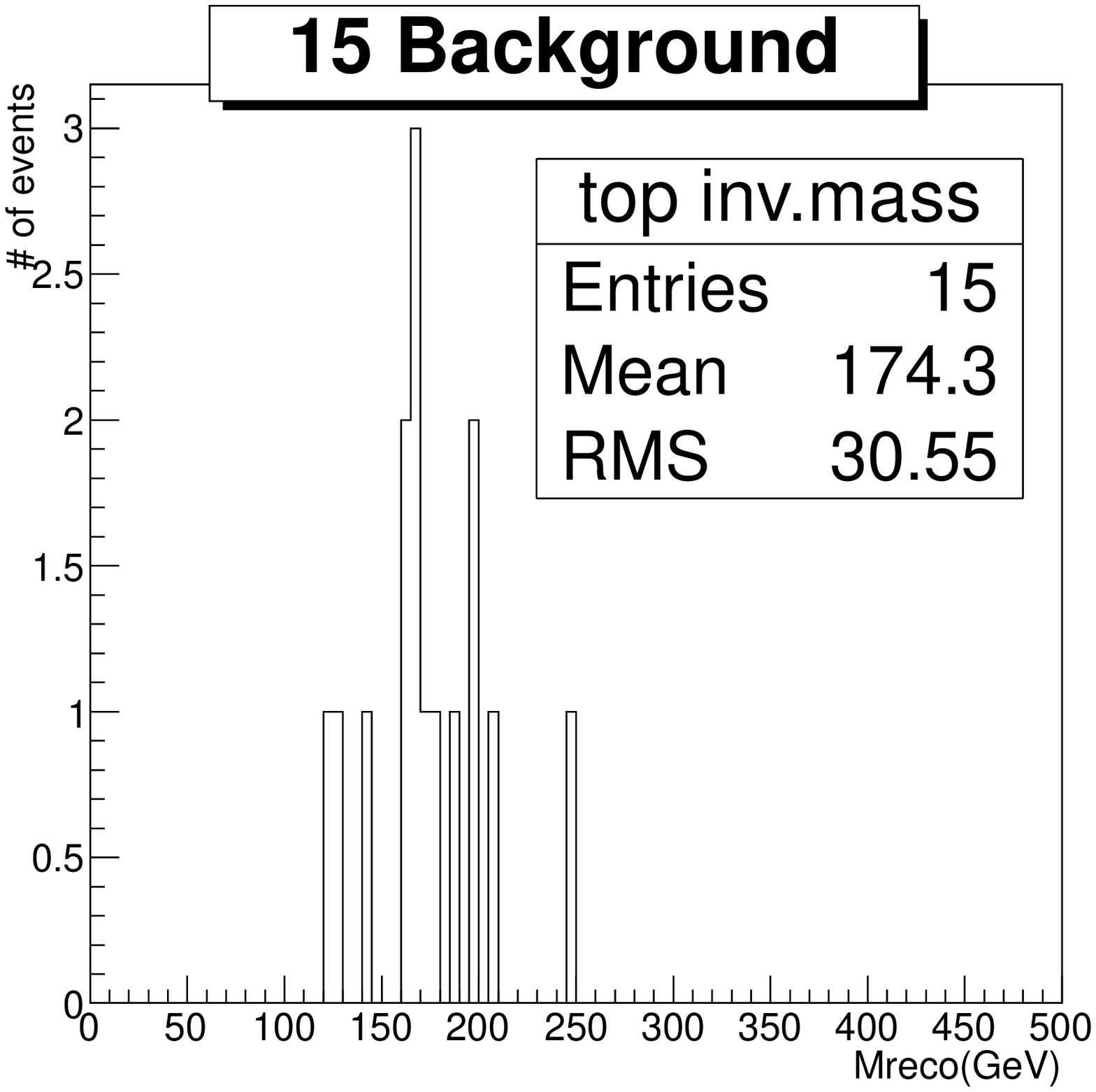}

\caption{Invariant mass histograms obtained from cut-based analysis for various
event samples with decreasing S/B ratio and an equal signal mass of
500 GeV.\label{fig:500gevcbresults}}
\end{figure}

\end{quotation}
This procedure was also tested with other $u{}_{4}$ masses, namely
400 and 600 GeV. The reconstructed invariant mass histograms for these
input masses are shown in Fig. \ref{fig:400gevcbresults} and Fig.
\ref{fig:600gevcbresults} .
\begin{quotation}
\centering

\begin{figure}[H]
\includegraphics[scale=0.26]{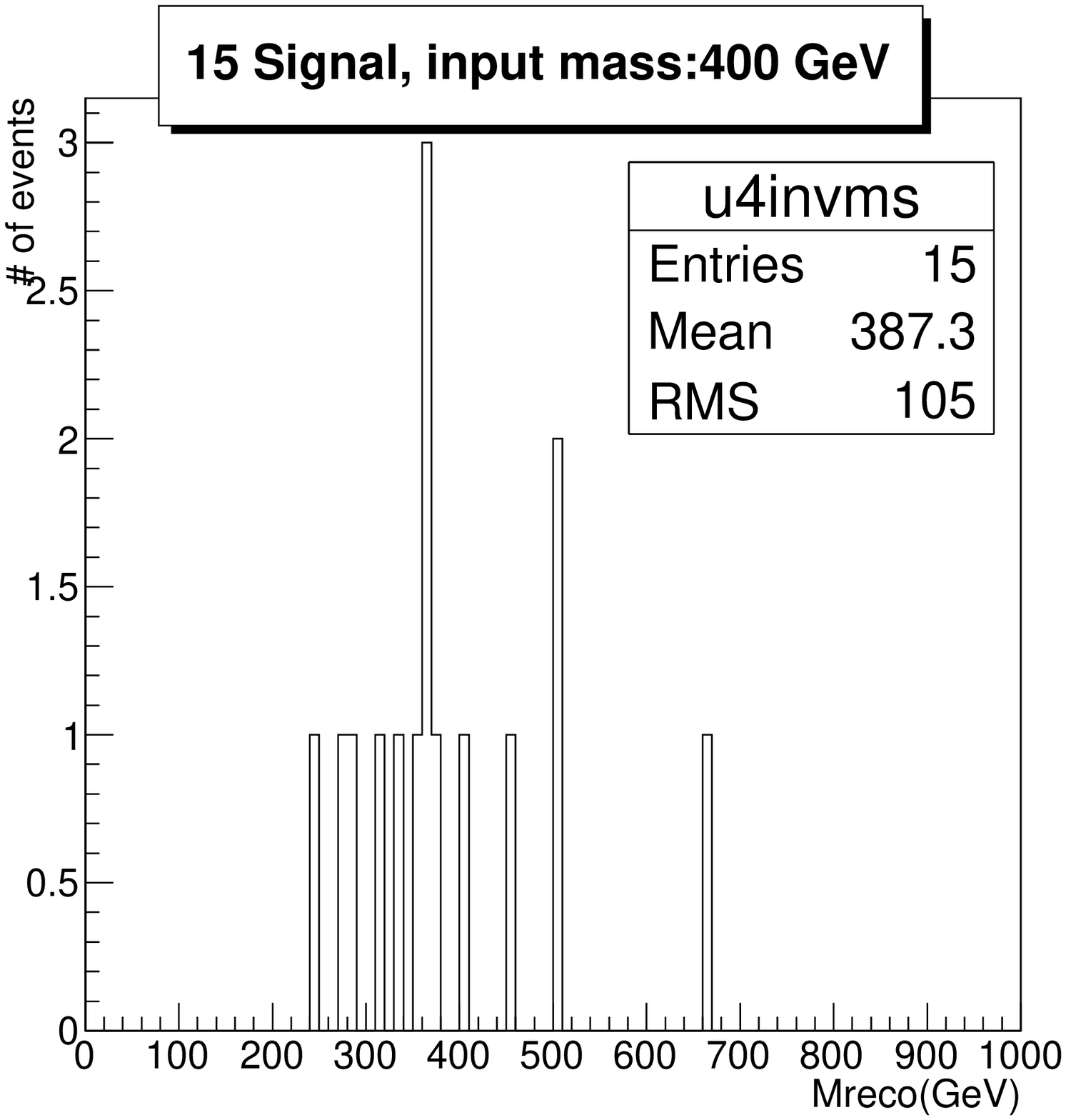}\includegraphics[scale=0.26]{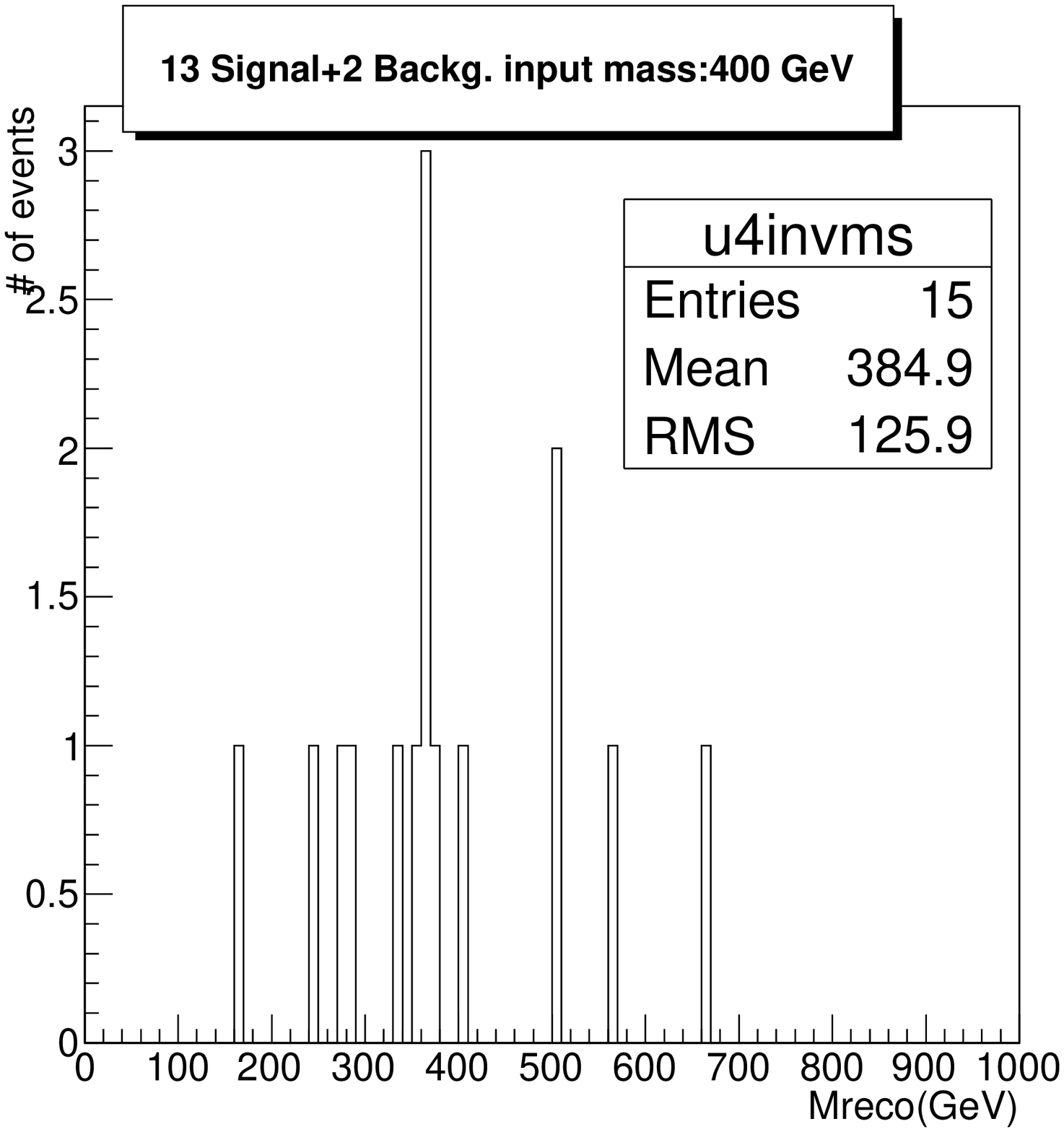}\includegraphics[scale=0.26]{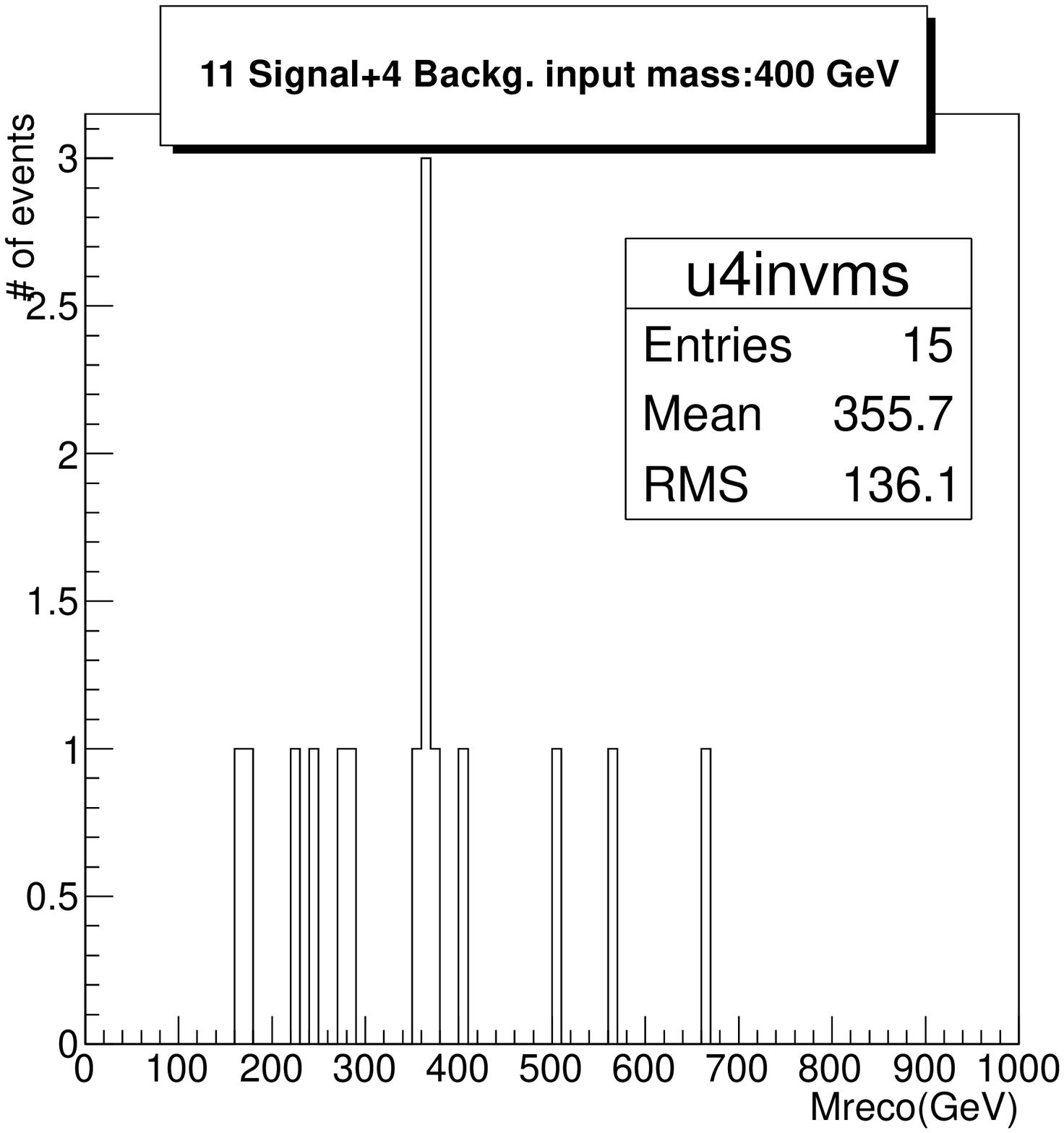}

\includegraphics[scale=0.26]{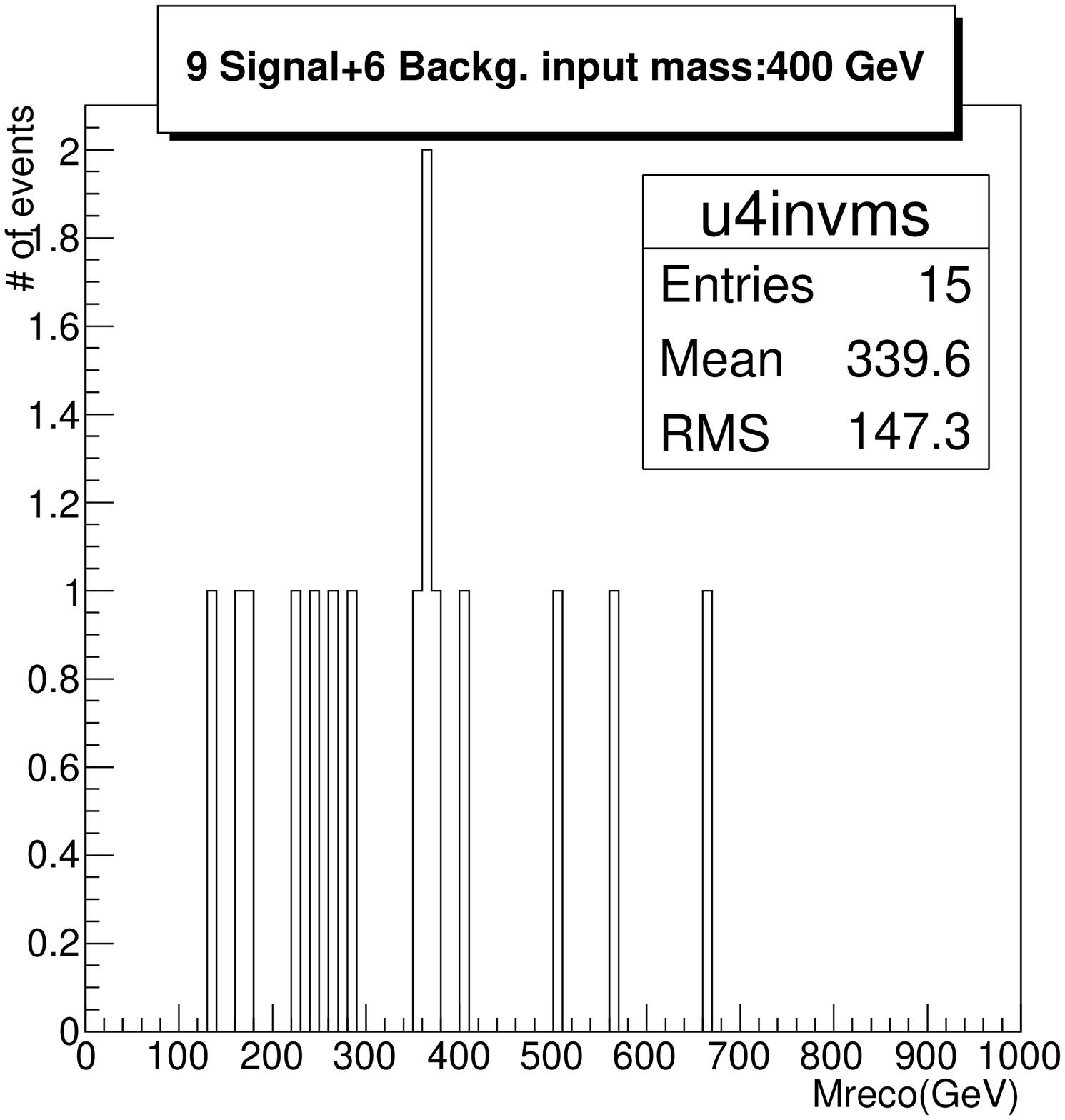}\includegraphics[scale=0.26]{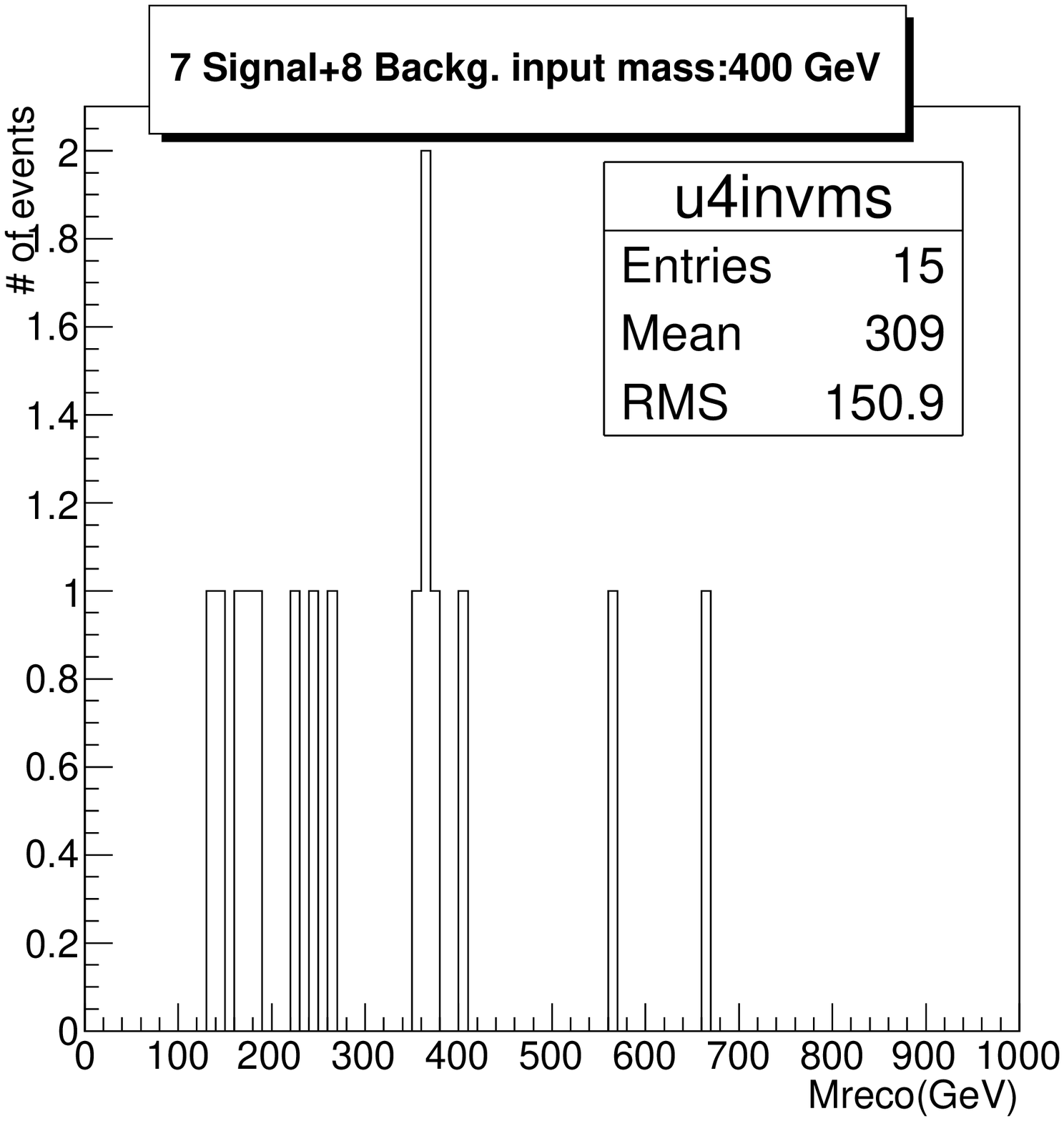}\includegraphics[scale=0.26]{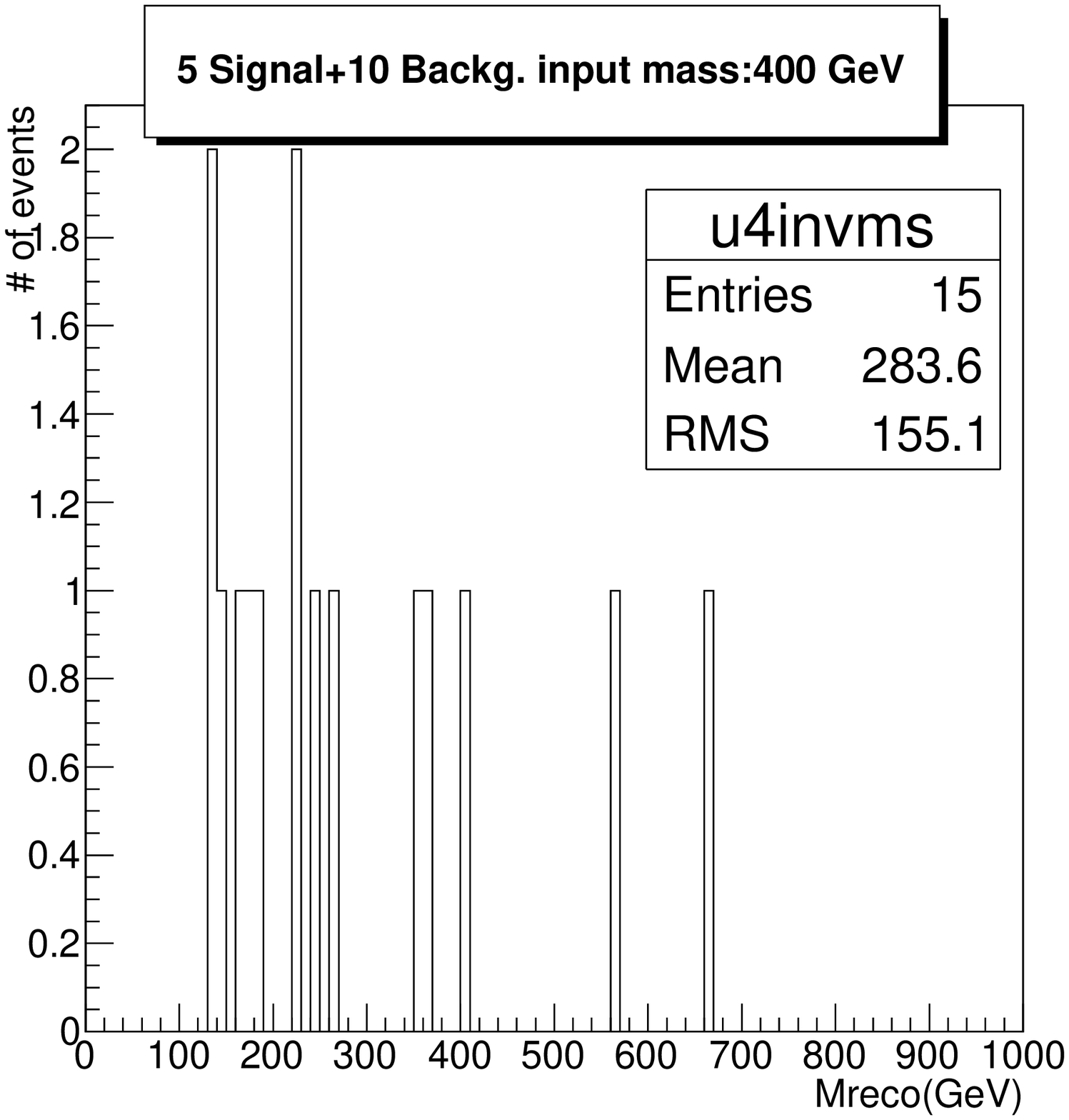}

\qquad{}\qquad{}\qquad{}\includegraphics[scale=0.26]{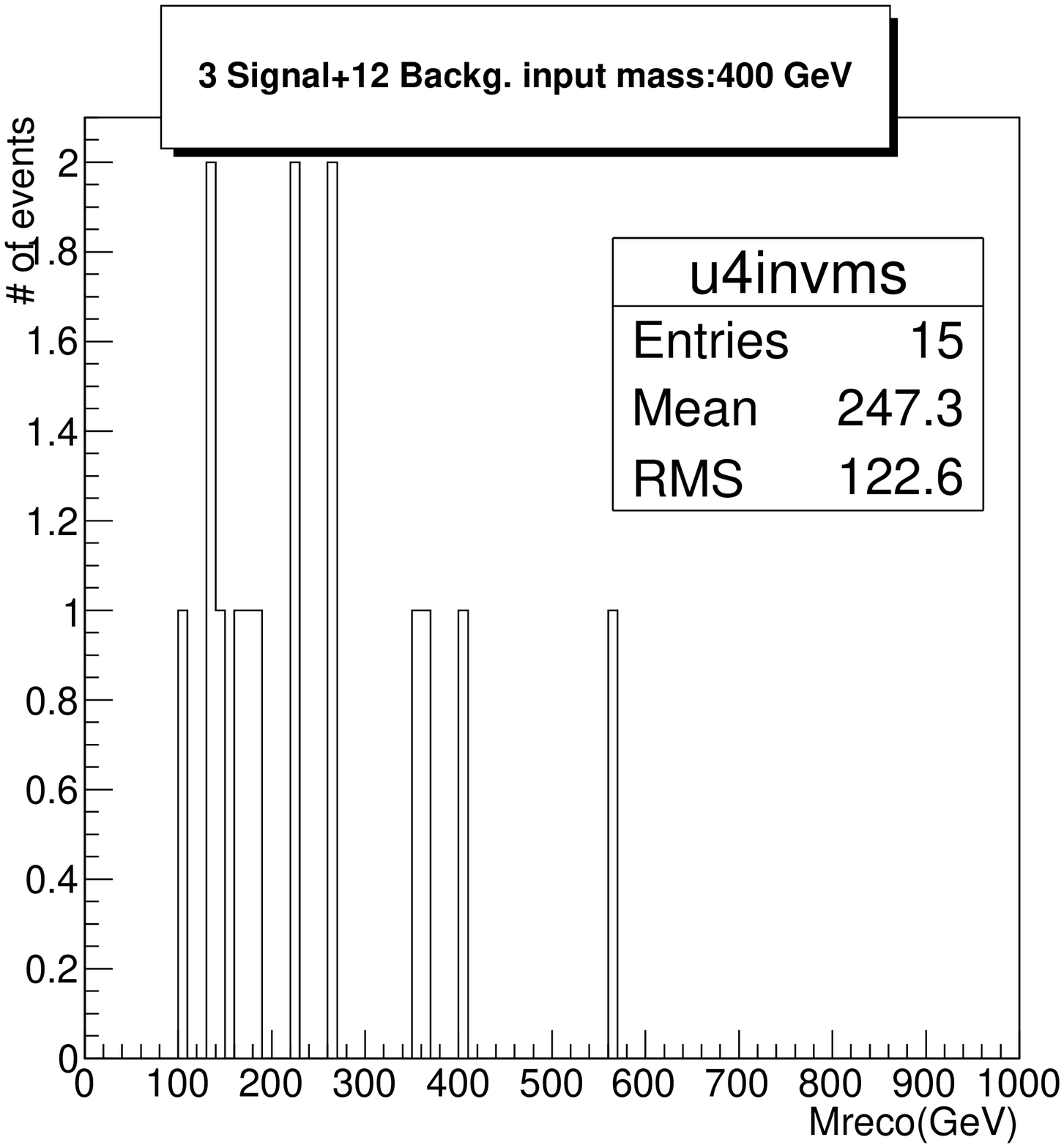}\ \includegraphics[scale=0.28]{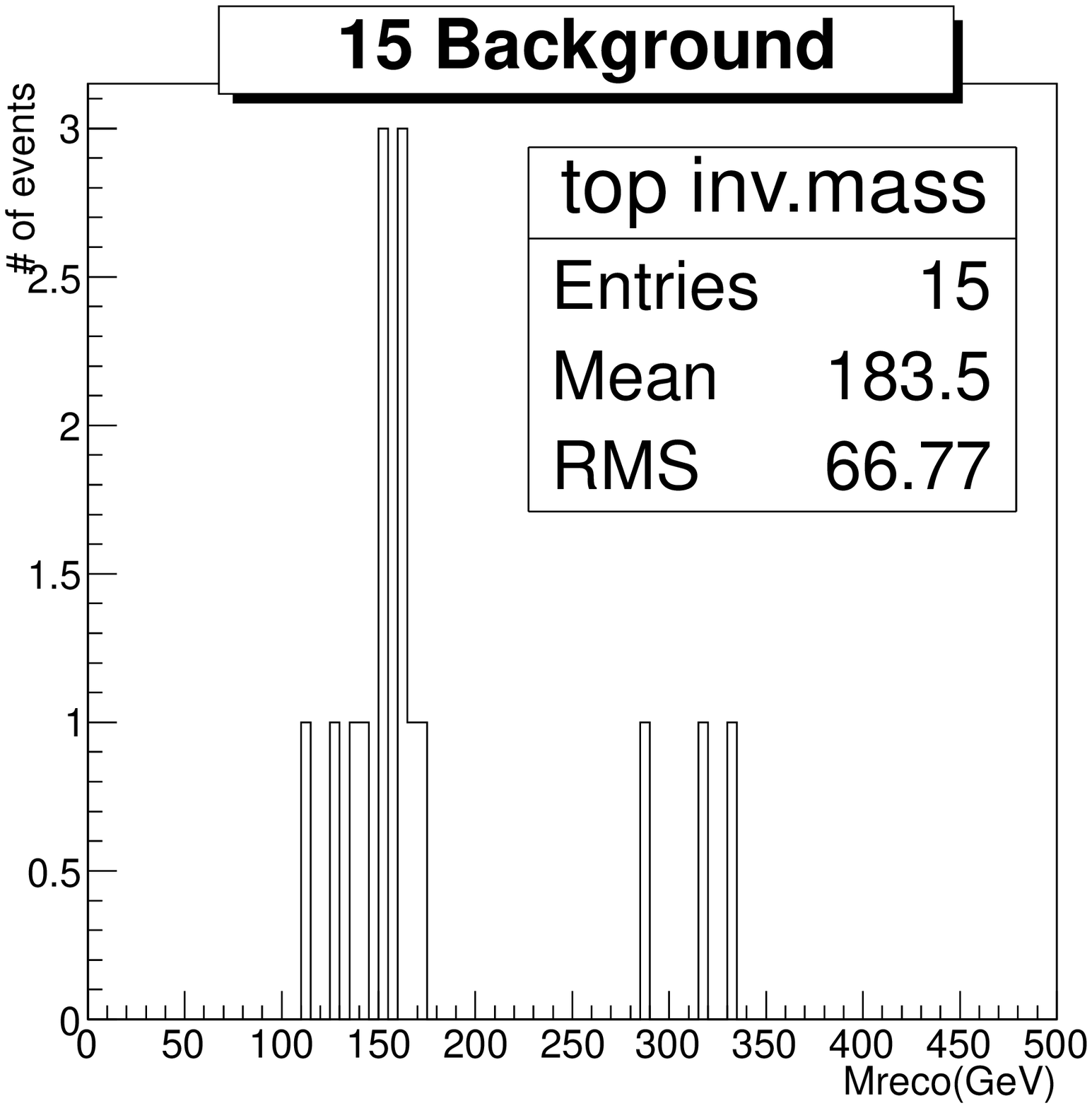}\caption{The same as Fig. \ref{fig:500gevcbresults} but for \textit{m$_{u_{4}}$=}
400 GeV.\label{fig:400gevcbresults}}
\end{figure}

\begin{figure}[H]
\includegraphics[scale=0.26]{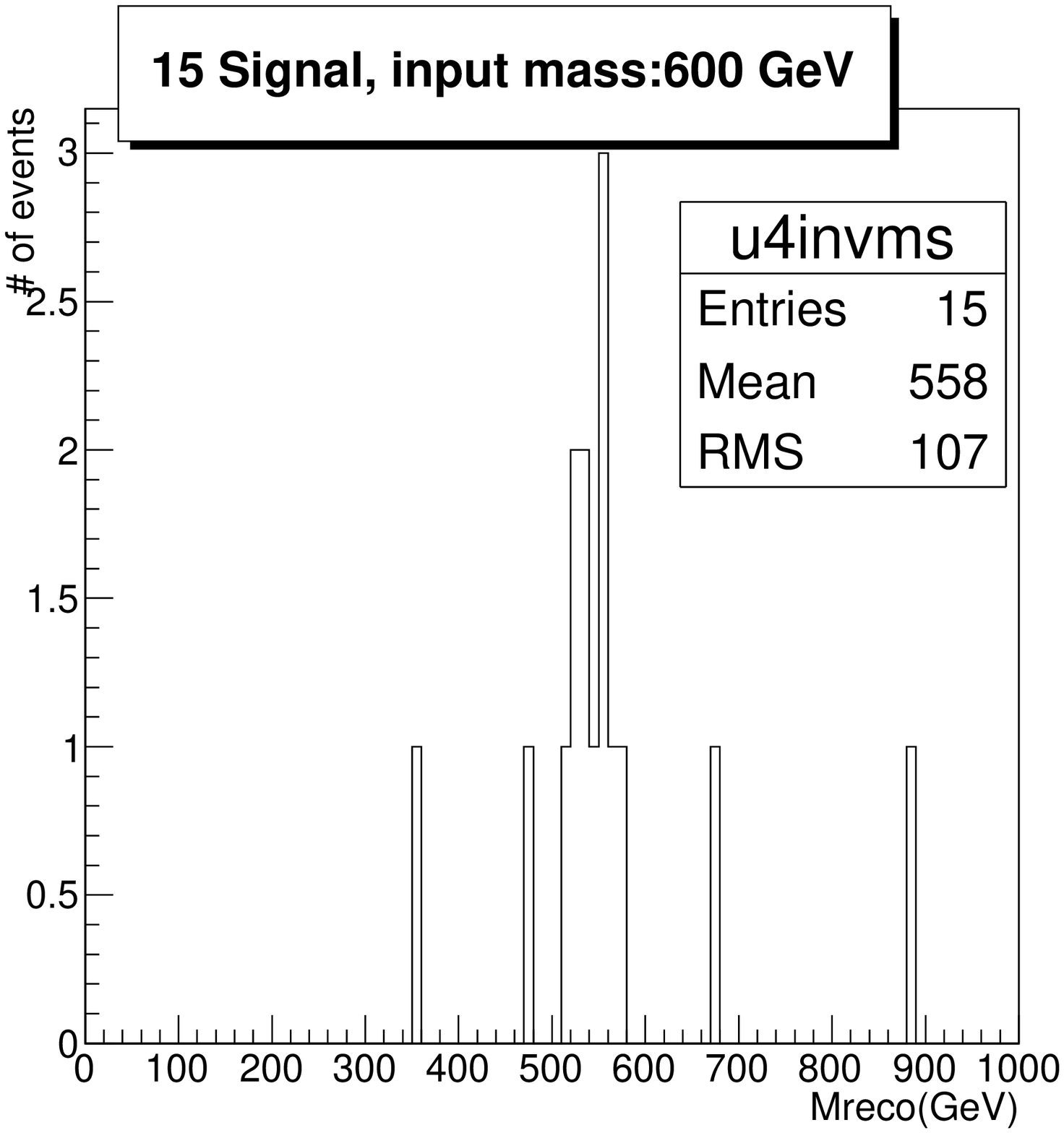}\includegraphics[scale=0.26]{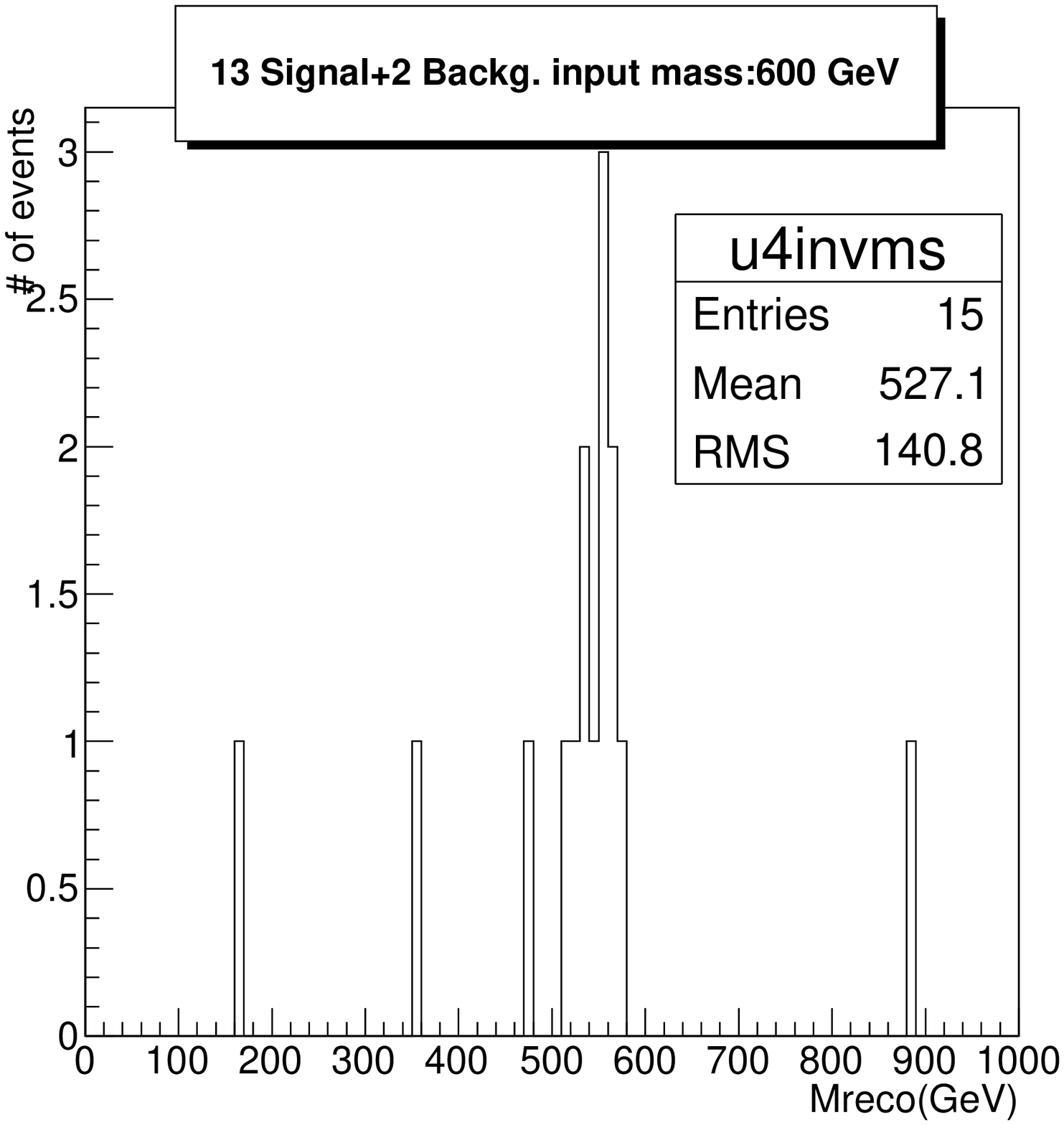}\includegraphics[scale=0.26]{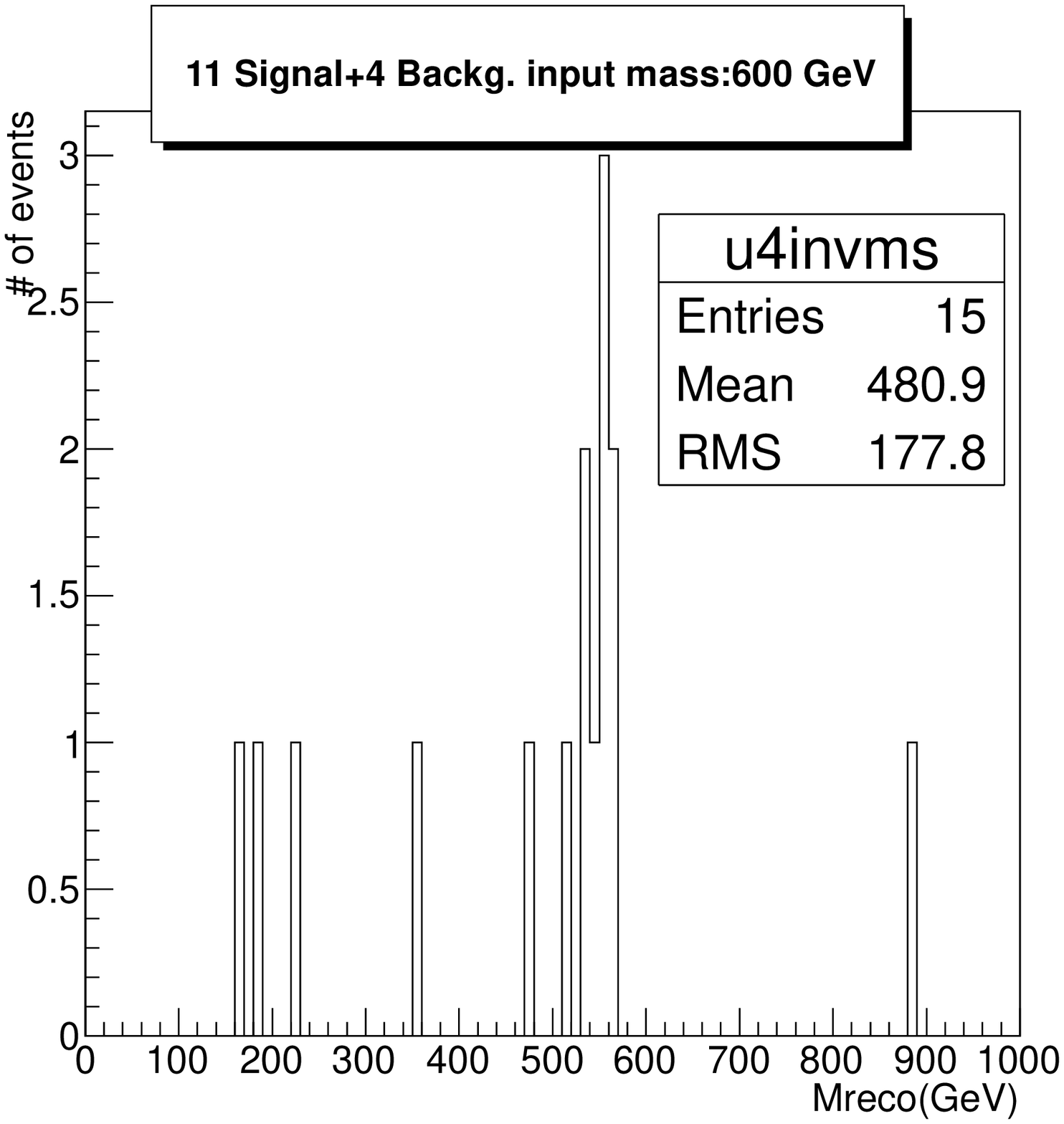}

\includegraphics[scale=0.26]{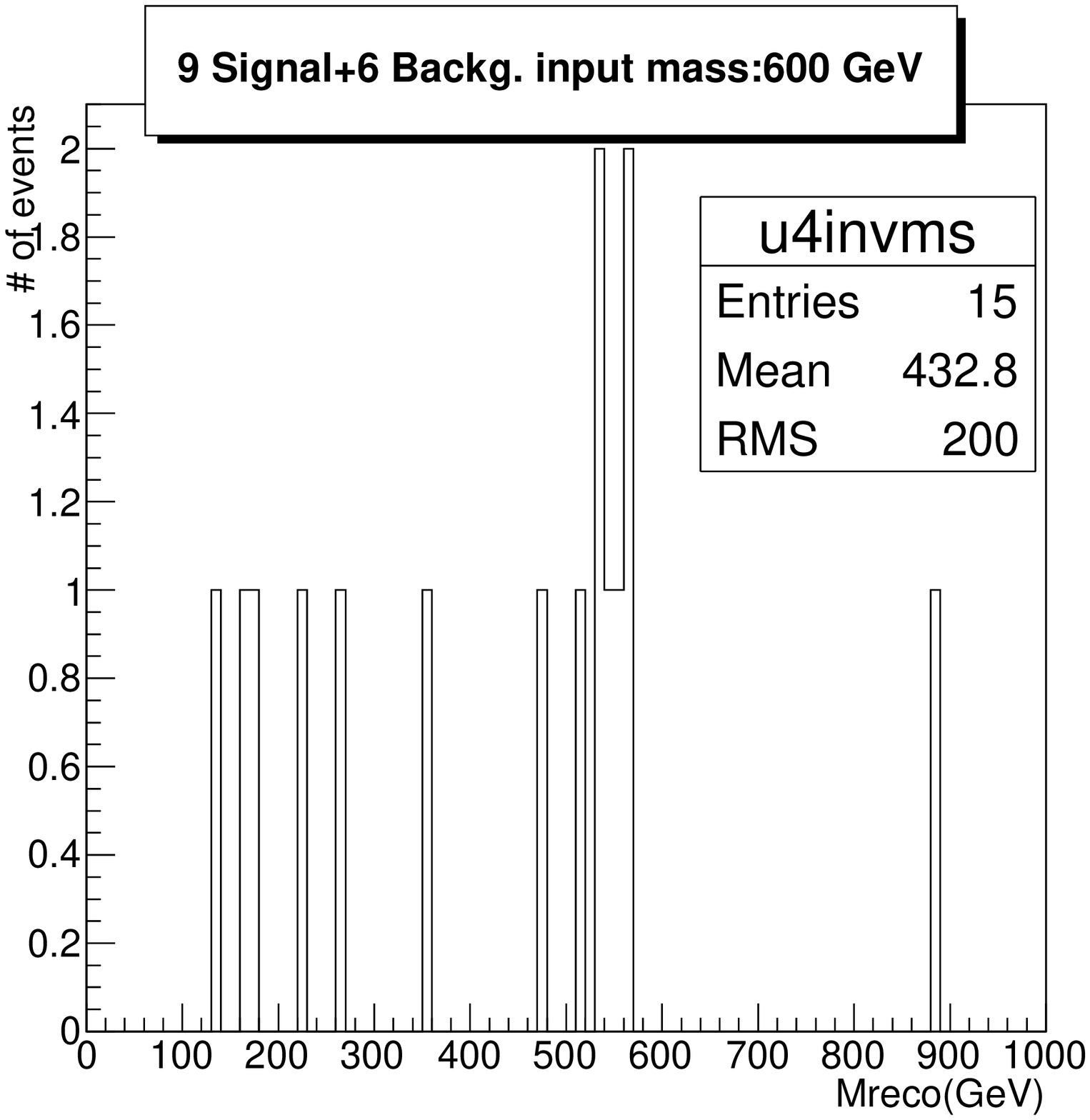}\includegraphics[scale=0.26]{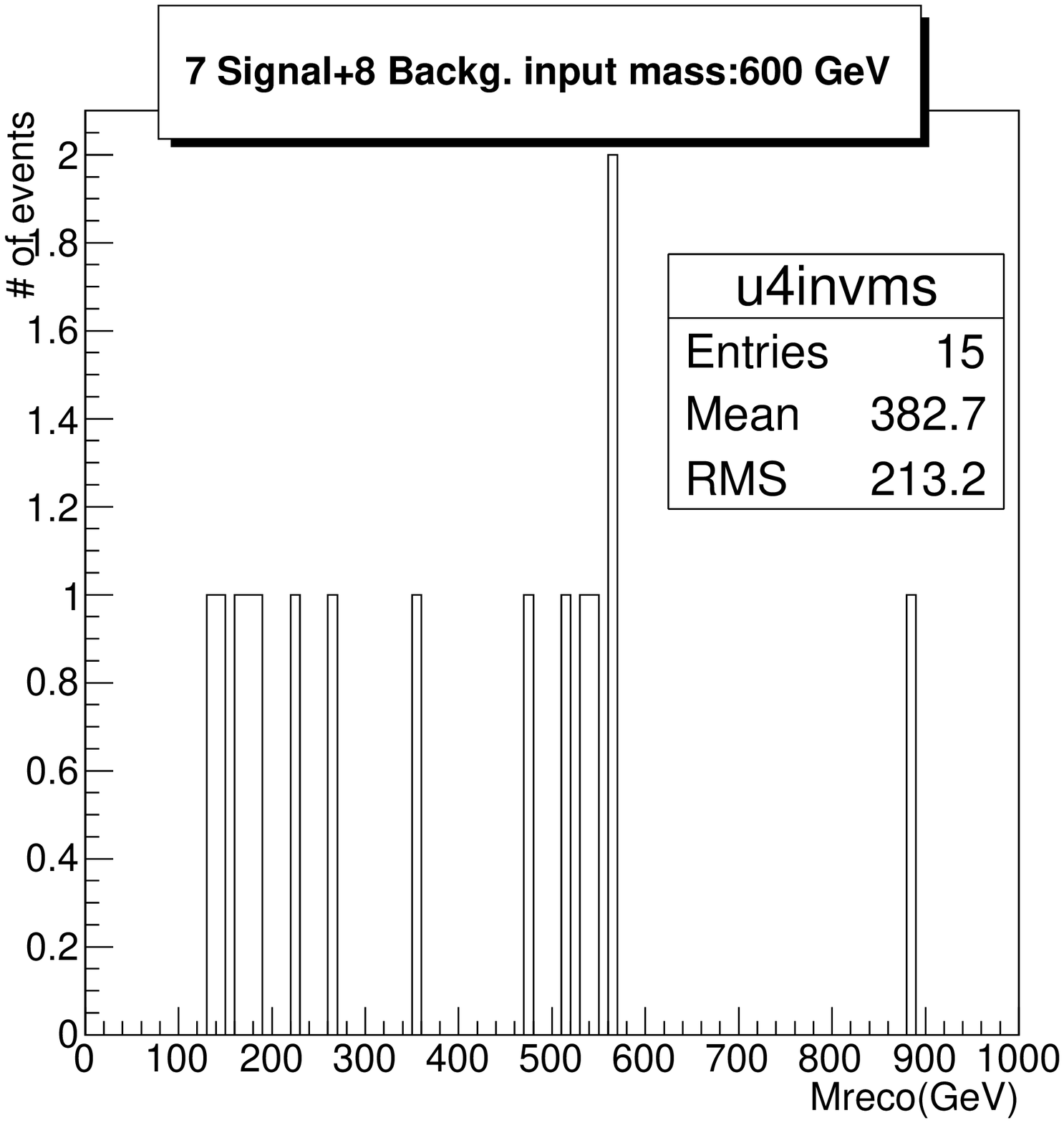}\includegraphics[scale=0.26]{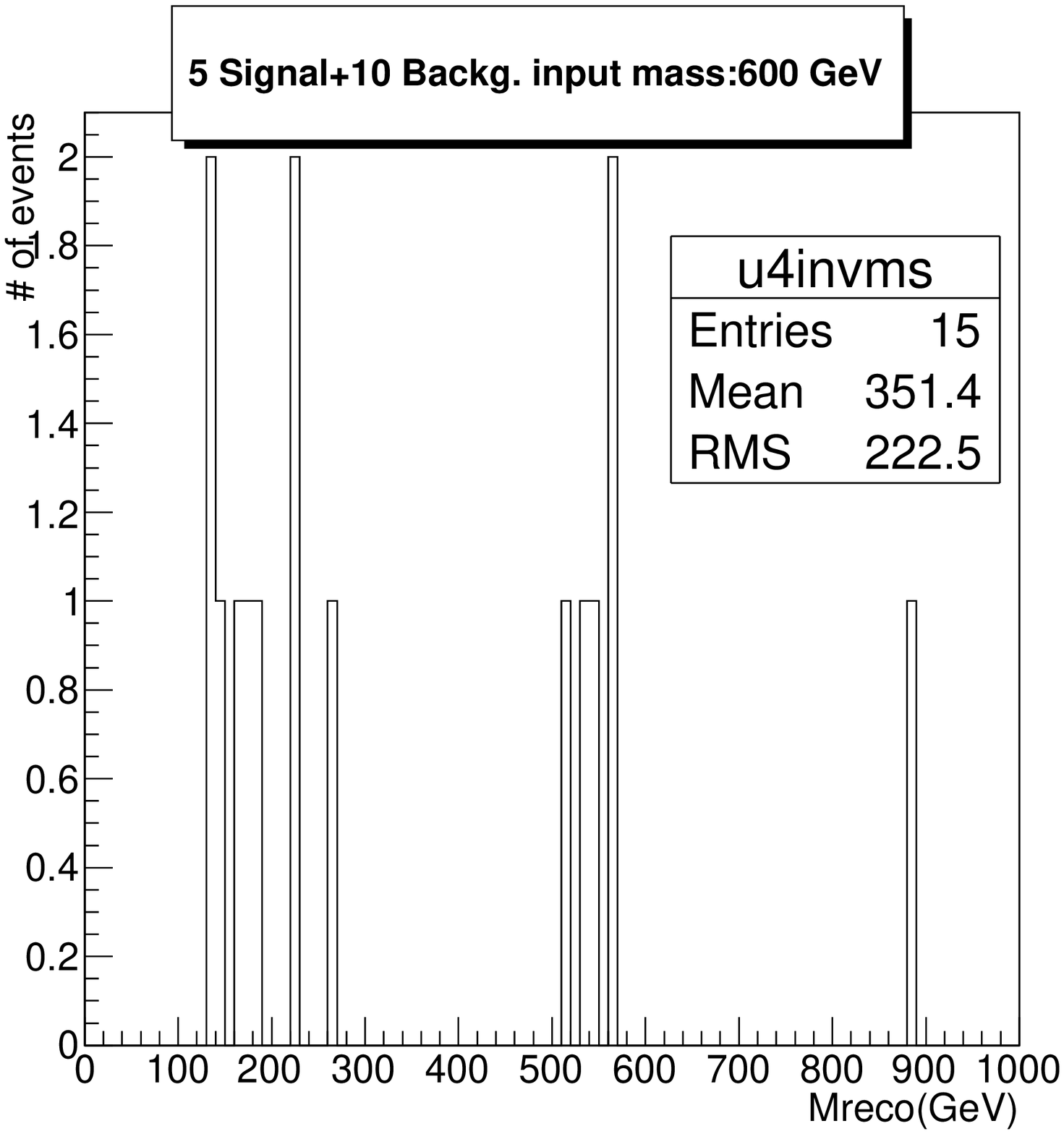}

\qquad{}\qquad{}\qquad{}\includegraphics[scale=0.26]{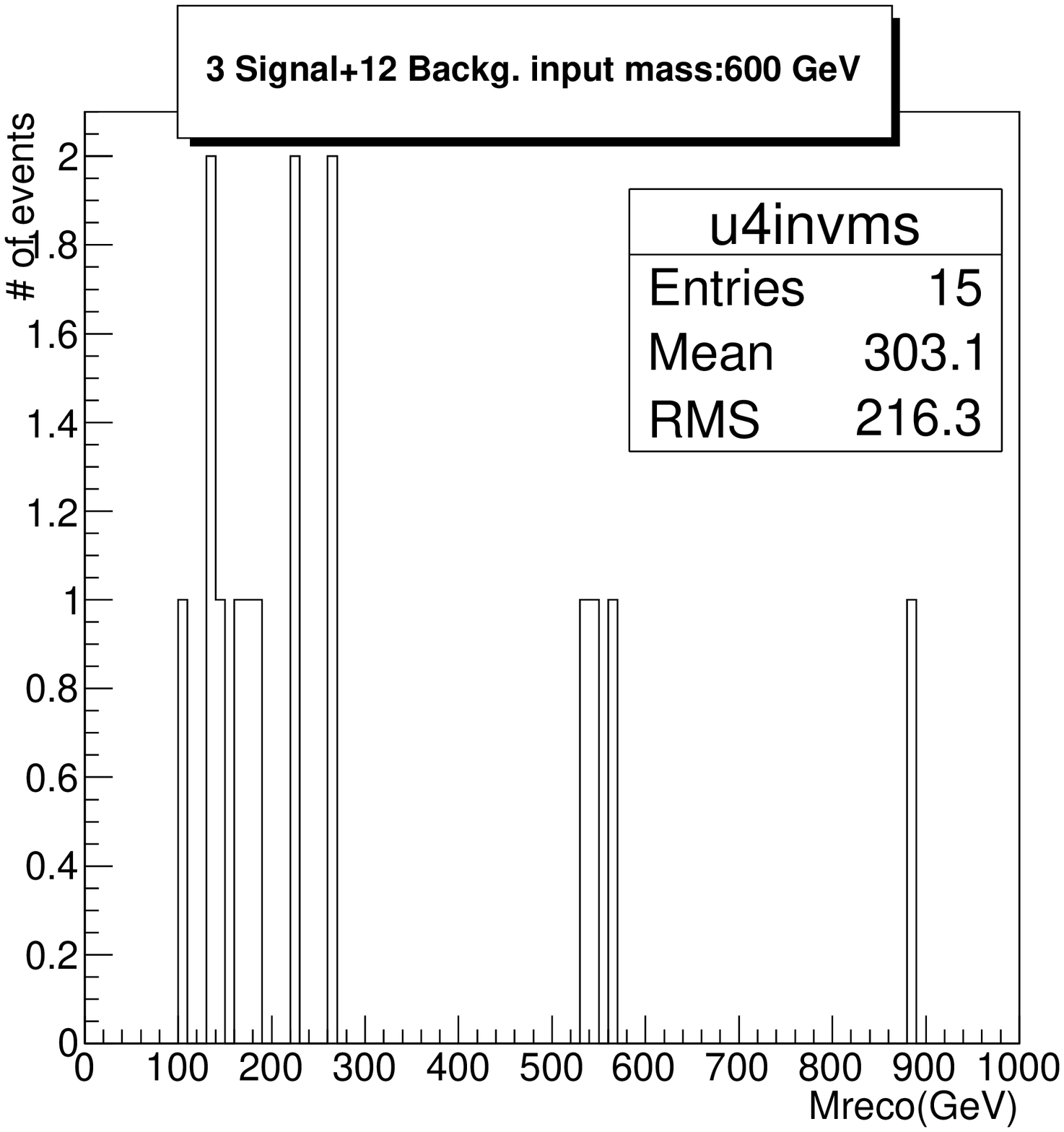}~\includegraphics[scale=0.28]{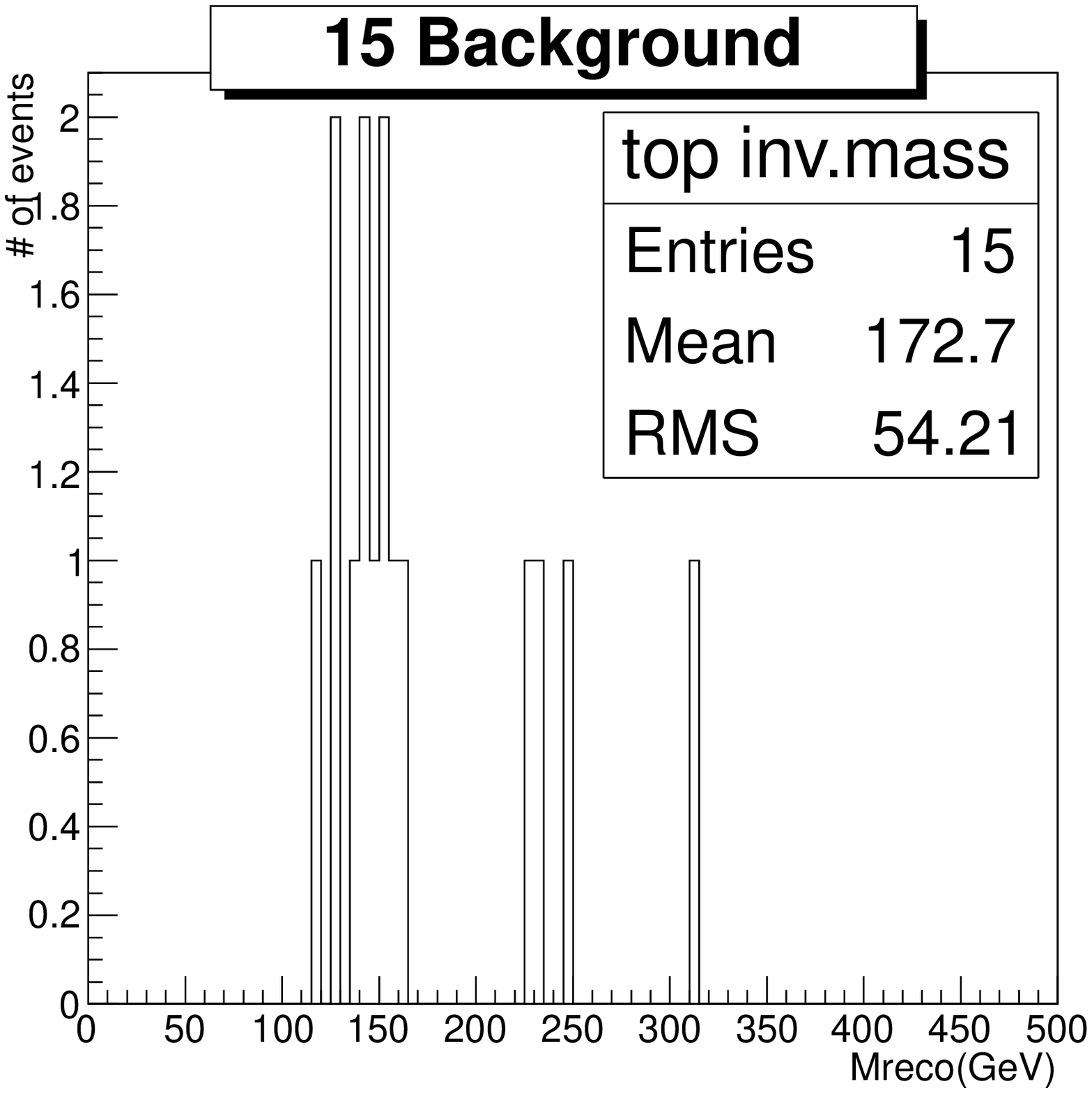}\caption{The same as Fig. \ref{fig:500gevcbresults} but for \textit{m$_{u_{4}}$=}
600 GeV.\label{fig:600gevcbresults}}
\end{figure}

\end{quotation}
\begin{flushleft}
The input masses and the reconstructed masses using cut-based technique
for the final states with different S/B ratios are shown in the Table
\ref{tab:cbasedtablo} .
\par\end{flushleft}

\begin{table}[H]
\caption{Invariant mass values extracted from the cut-based analysis for various
samples which have different S/B ratios and different input values.\label{tab:cbasedtablo}}
\end{table}

\noindent %
\begin{tabular}{|l|c|c|c|}
\hline 
\multirow{2}{*}{Event sample} & \multicolumn{3}{c|}{Output \textit{u$_{4}$} masses for }\tabularnewline
\cline{2-4} 
 & input mass= 400 GeV & input mass= 500 GeV & input mass= 600 GeV\tabularnewline
\hline 
15 signal & \multicolumn{1}{c|}{387.3 $\pm$ 105} & 450.7 $\pm$ 113.5 & 558 $\pm$ 107\tabularnewline
\hline 
13 signal + 2 backg. & 384.9 $\pm$ 125.9 & 446.3 $\pm$ 125.4 & 527.1 $\pm$ 140.8\tabularnewline
\hline 
11 signal + 4 backg. & 355.7 $\pm$ 136.1 & 422.3 $\pm$ 148.3 & 480.9 $\pm$ 177.8\tabularnewline
\hline 
9 signal + 6 backg. & 339.6 $\pm$ 147.3 & 387.4 $\pm$ 166.6 & 432.8 $\pm$ 200\tabularnewline
\hline 
7 signal + 8 backg. & 309 $\pm$ 150.9 & 349.3 $\pm$ 179.7 & 382.7 $\pm$ 213.2\tabularnewline
\hline 
5 signal + 10 backg. & 283.6 $\pm$ 155.1 & 326.1 $\pm$ 188.5 & 351.4 $\pm$ 222.5 \tabularnewline
\hline 
3 signal + 12 backg. & 247.3 $\pm$ 122.6 & 289.7 $\pm$ 187.8 & 303.1 $\pm$ 216.3\tabularnewline
\hline 
0 signal + 15 backg. & 183.5 $\pm$ 66.77  & 174.3 $\pm$ 30.55  & 172.7 $\pm$ 54.21 \tabularnewline
\hline 
\end{tabular}

\ 

The error values shown here indicate only statistical errors (systematics
effects are not considered at this stage). One can see from Table
\ref{tab:cbasedtablo} that even in the case of pure signal sample,
the deviation from input values is large and the most correct result
is obtained for 400 GeV input mass. These huge uncertainties seen
here came from the low statistics. The second interesting point is
that, the samples including mostly background events also give new
quark mass estimations around \textit{u$_{4}$ }input mass instead
of top mass, therefore this approach is relatively useless for discriminating
signal and background events especially with low statistics.

\subsection{Matrix Element Method Analysis:}

This method relies on the correct calculation of the weights in Eq.
\ref{eq:olasilik}. To ensure their correct computation, MadWeight
{[}11{]}, which was developed by the MadGraph Team, has been used.
MadWeight is a phase space generator which takes lhco files {[}20{]}
and processes information with data cards and returns likelihood values
for the parameter of interest. 

In this part, event files for 15 signal, 13 signal+2 background, 11
signal+4 background and so forth are used in MadWeight to estimate
the signal mass for three input \textit{u$_{4}$} masses: 400, 500
and 600 GeV. A sample of N = 15 events are processed through MadWeight
for the evaluation of the weights. The mass of the $u$$_{4}$ quark
is extracted through the minimization of \textminus{}\textit{ln}(\textit{L}(\textit{m$_{u_{4}}$}))
with respect to the \textit{m$_{u_{4}}$}.

In this note, the default transfer function in MadWeight has been
used. In the default TF set, the jet energy is parametrized by a double
gaussian, and all other quantities such as the angles of visible particles
and the energy of leptons are assumed to be well measured. This means
that the corresponding transfer functions for lepton energies and
angles are given by delta functions. The transfer function associated
with a neutrino (MET) is taken to be one.

\begin{figure}[H]
\includegraphics[scale=0.26]{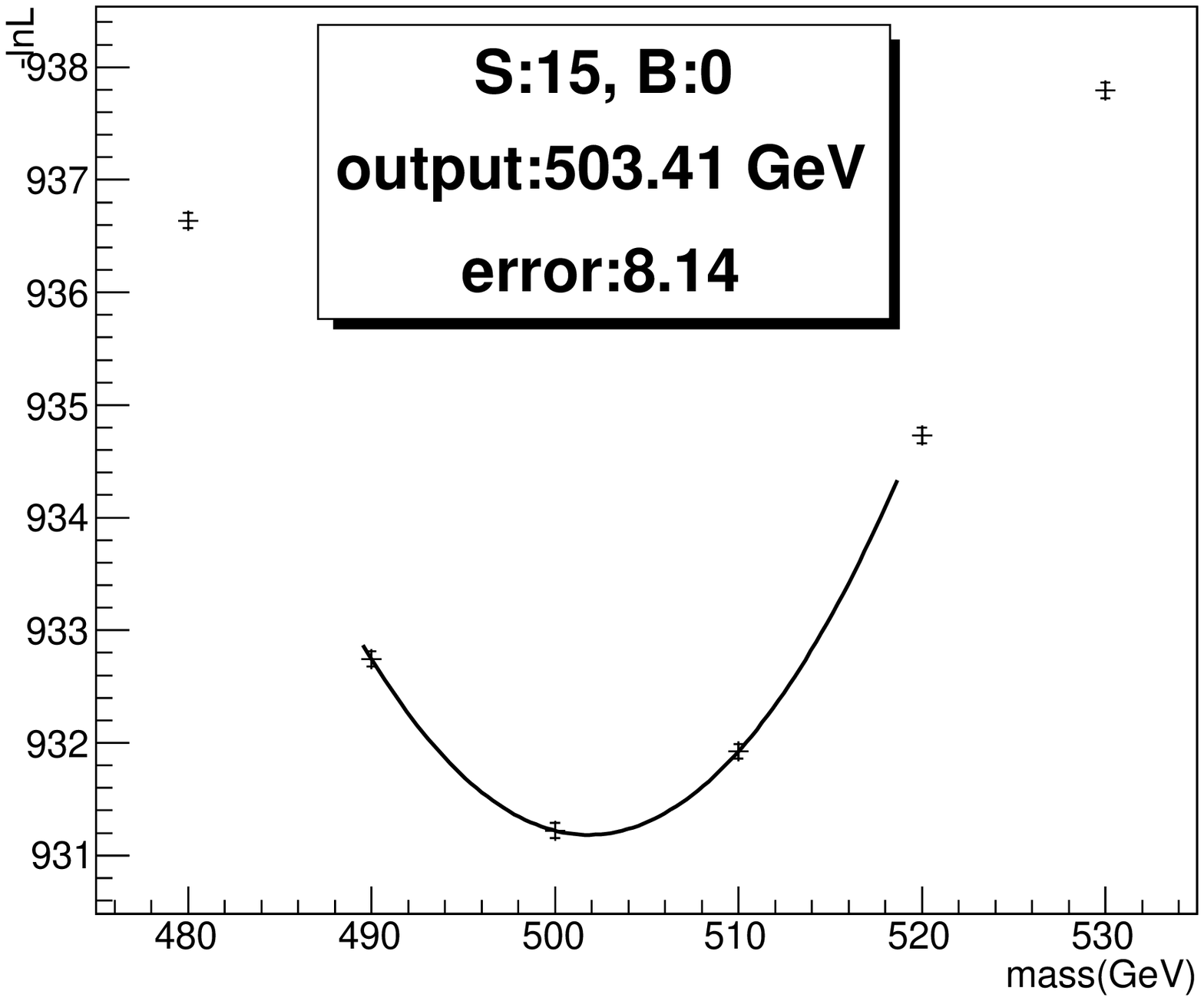}\includegraphics[scale=0.26]{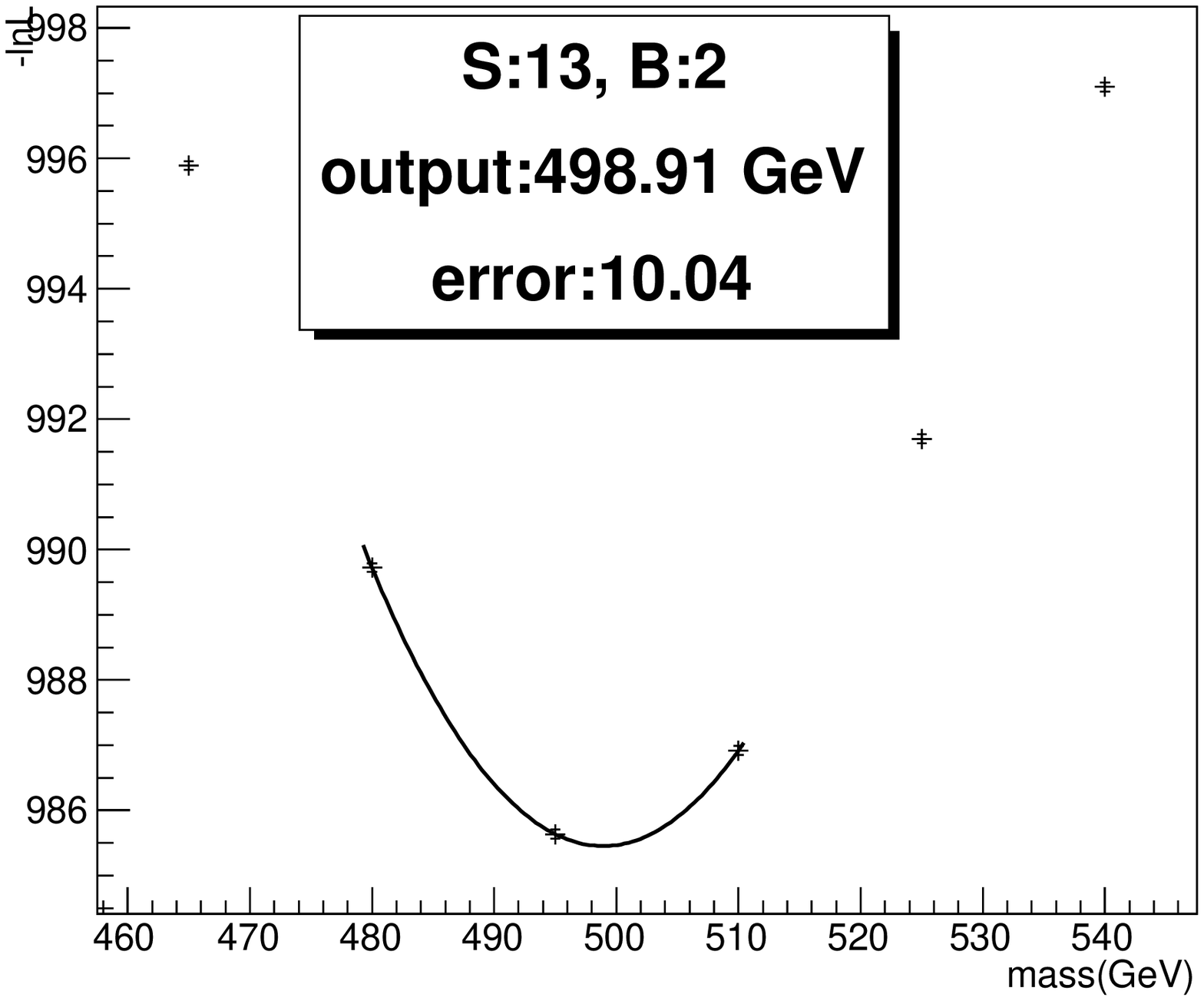}\includegraphics[scale=0.26]{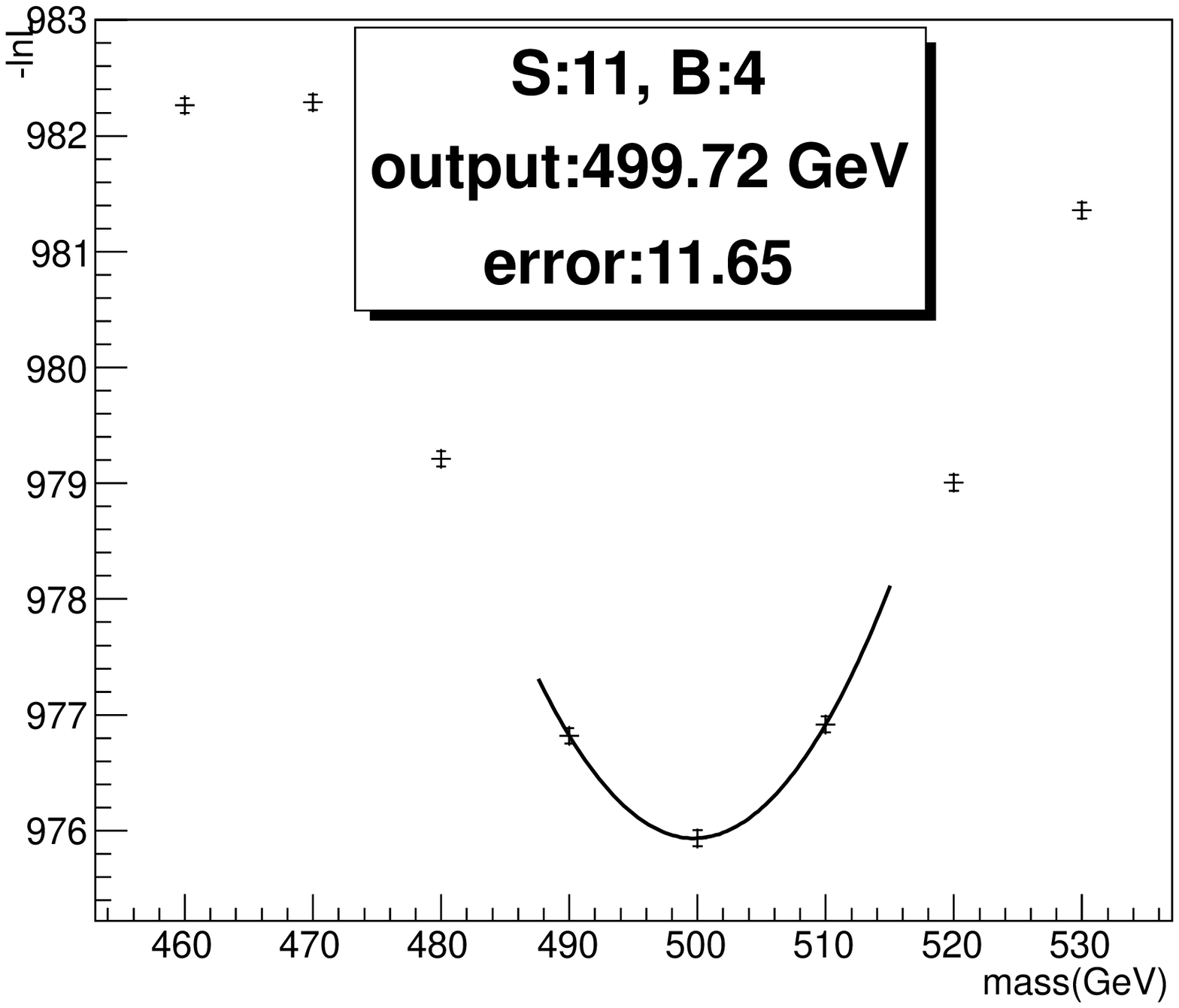}

\includegraphics[scale=0.26]{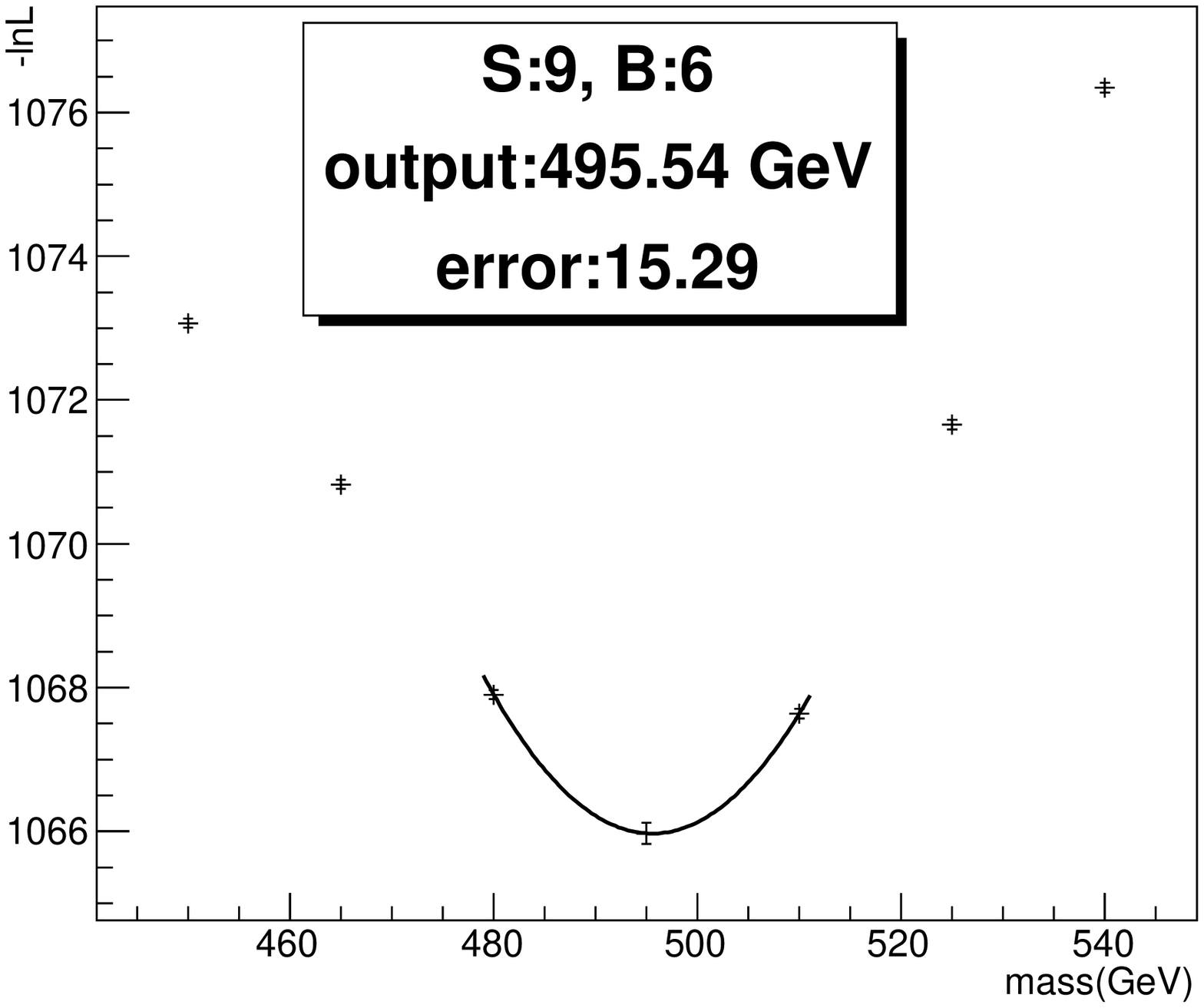}\includegraphics[scale=0.26]{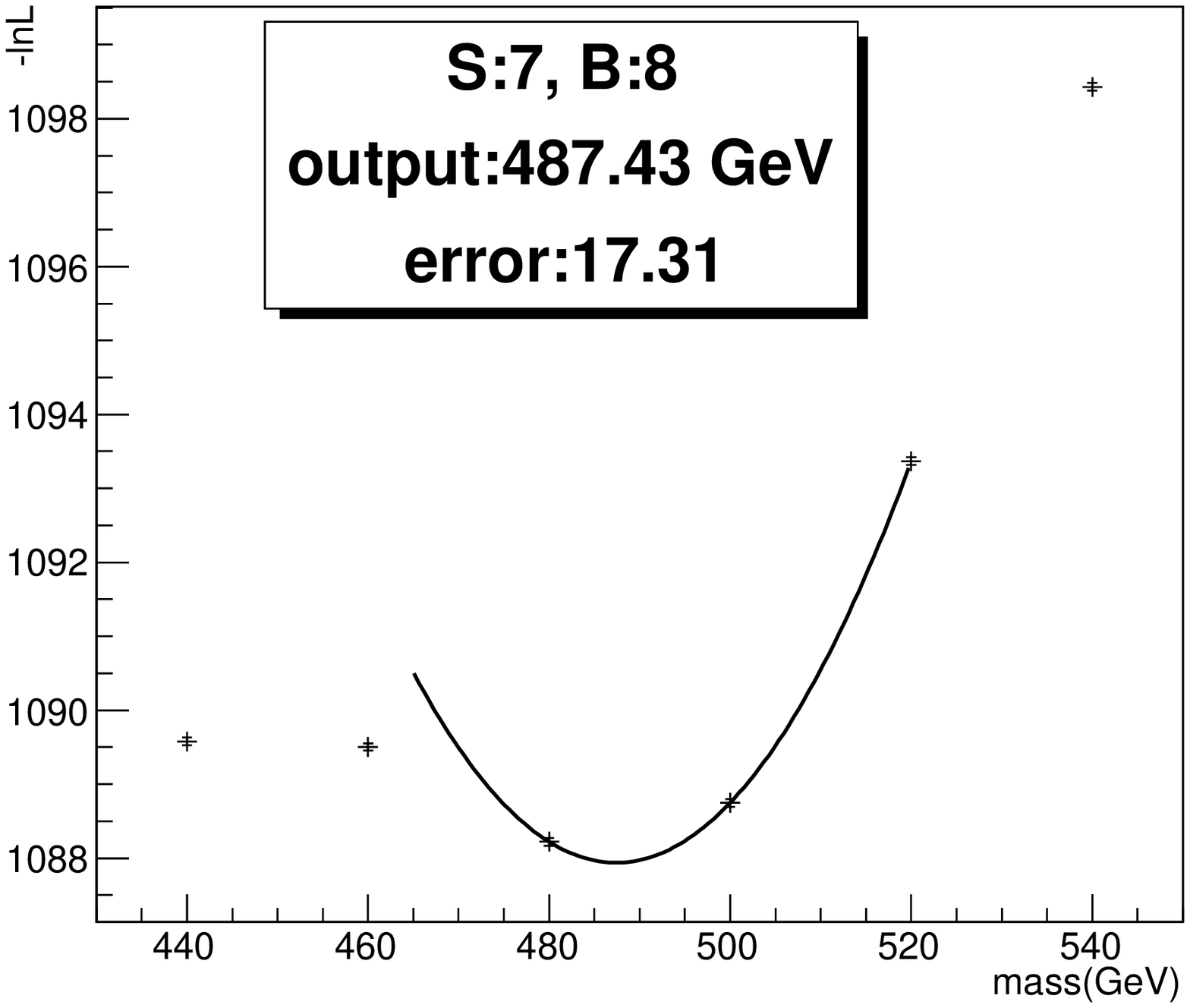}\includegraphics[width=5.2cm,height=4.4cm]{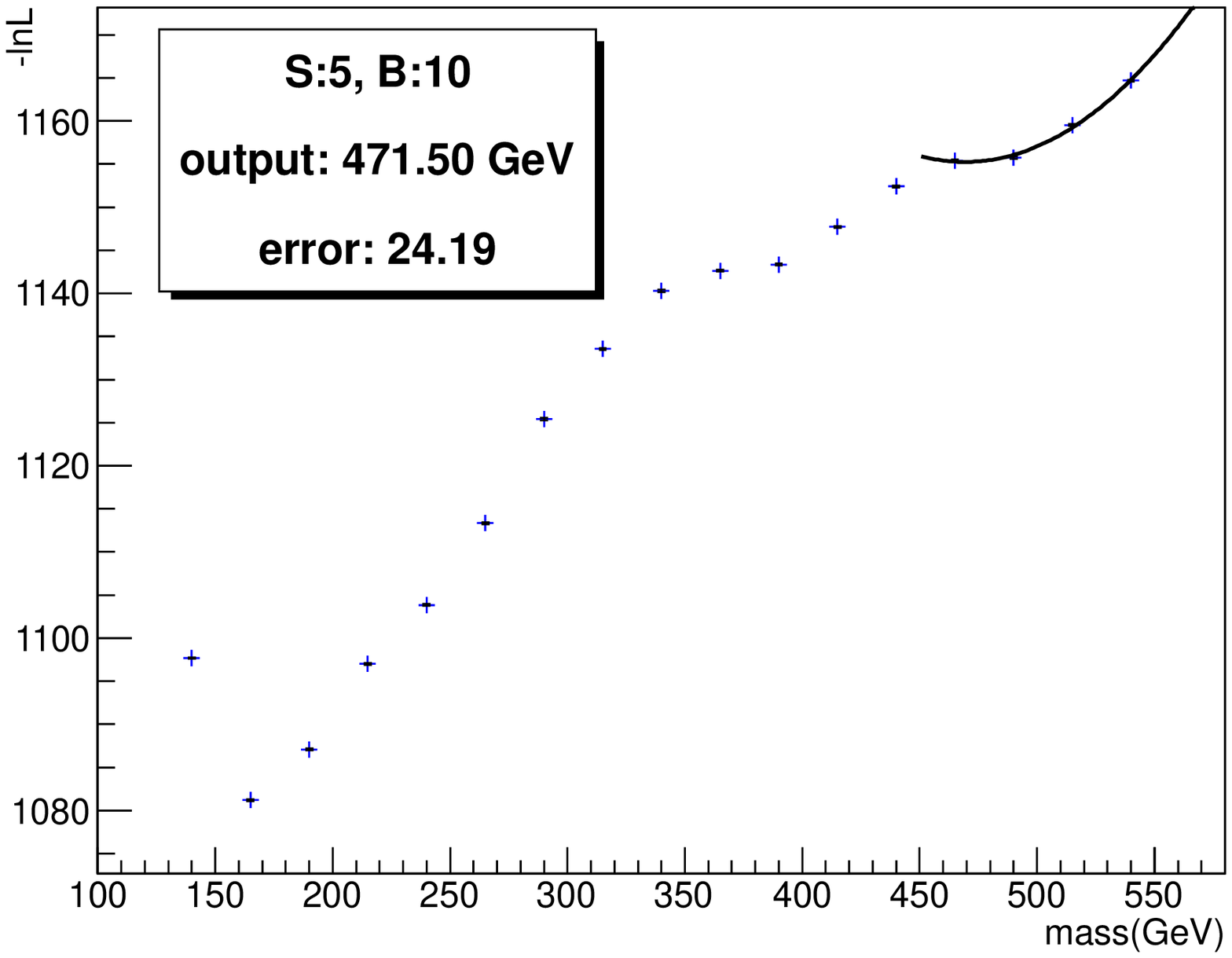}

\qquad{}\qquad{}\qquad{}\qquad{}\includegraphics[width=5.2cm,height=4.4cm]{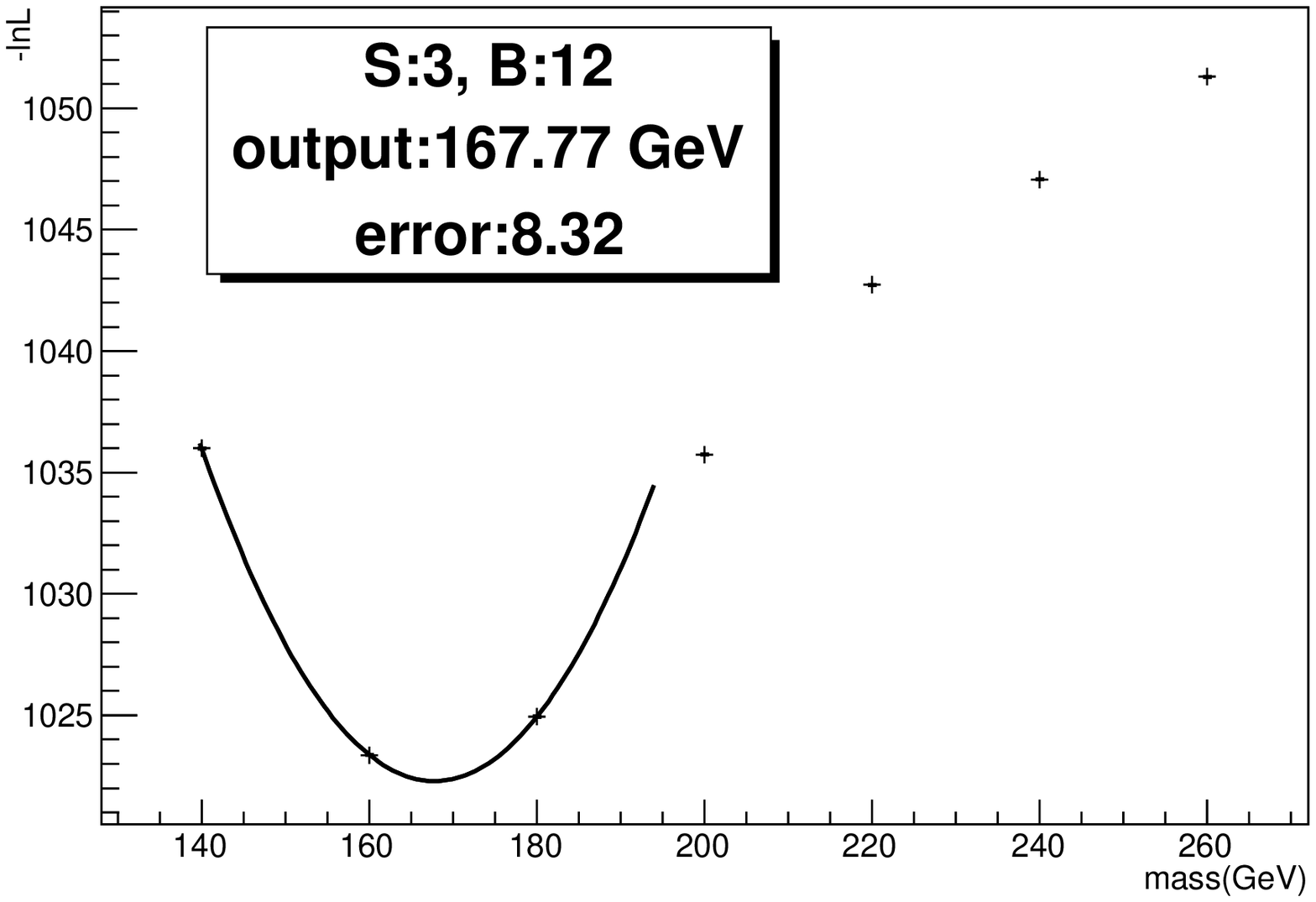}~\includegraphics[scale=0.28]{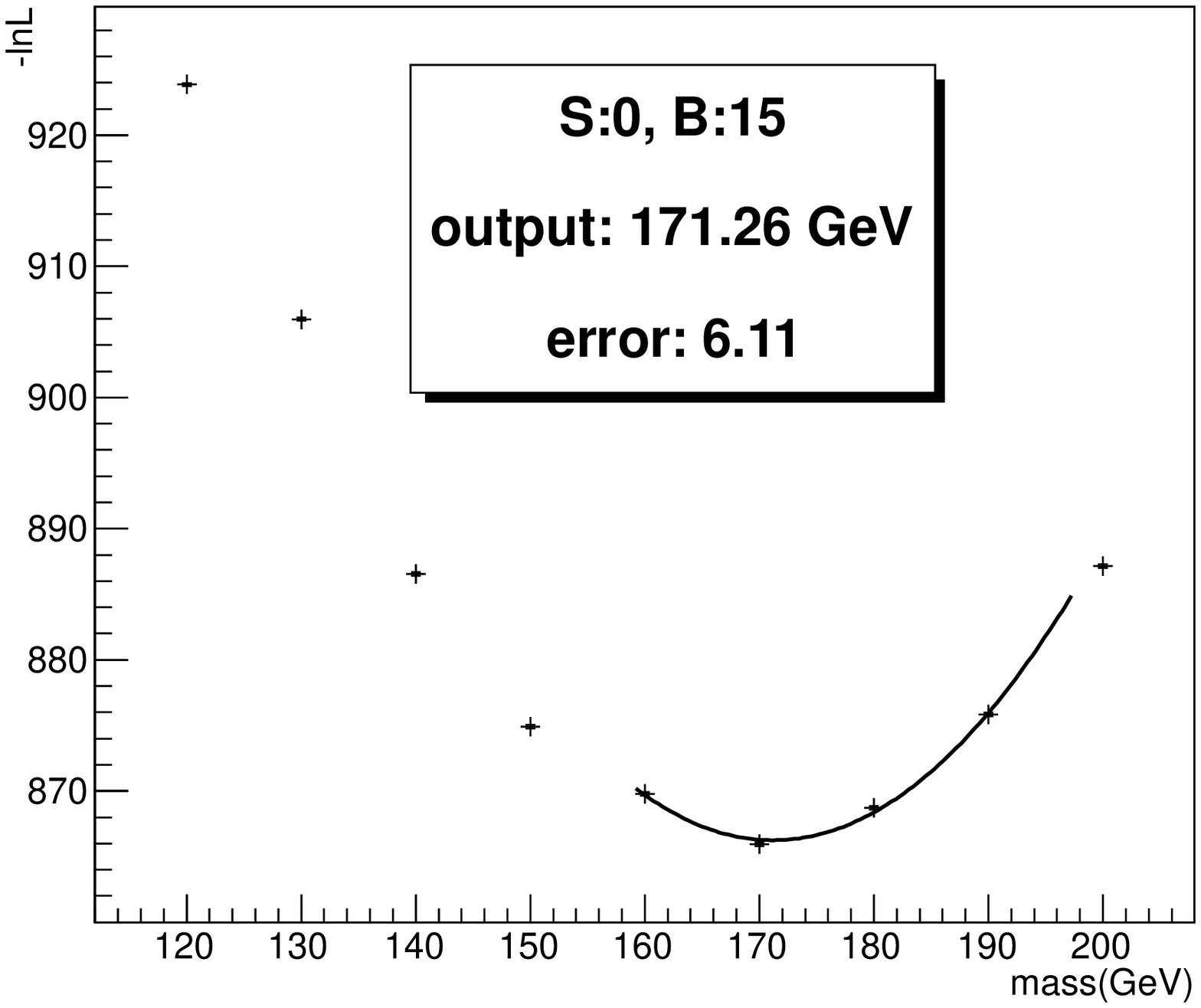}

\caption{Plots for likelihoods for samples of 15 events containing different
ratios of S/B generated with input mass of \textit{u$_{4}$} 500 GeV.
The mass value of \textit{u$_{4}$} has been extracted from the parabolic
curve fitting of the points around the minima.\label{fig:500gevmem plots}}
\end{figure}

As in the cut-base approach, the analysis started from event samples
which were generated with an input mass of 500 GeV. The likelihood
curves obtained for this mass with various signal and background samples
are shown in Fig. \ref{fig:500gevmem plots}.

Estimated \textit{u$_{4}$} masses and statistical errors are shown
in the legend box of each graph, except the last one, i.e. 3S plot
in which one finds 167.77 GeV. These estimations are extracted from
a parabolic curve fit to ($-lnL,Mass)$ points obtained from MadWeight.
Error values include both standard deviation of likelihoods, evaluated
via increasing the minimum likelihood value by 1/2, which corresponds
to a 1$\sigma$ deviation and also the errors originating from parabola
fitting. If a wide mass range is scanned, then two likelihood minima
are obtained $(top,u_{4})$ except the 3S12B case, where only one
value corresponding to the top quark mass is found. As seen in 5S10B
plot, there are two local minima between 350-550 GeV interval, and
the nearest one to \textit{u$_{4}$ }input mass is chosen.

\begin{figure}[H]
\includegraphics[scale=0.26]{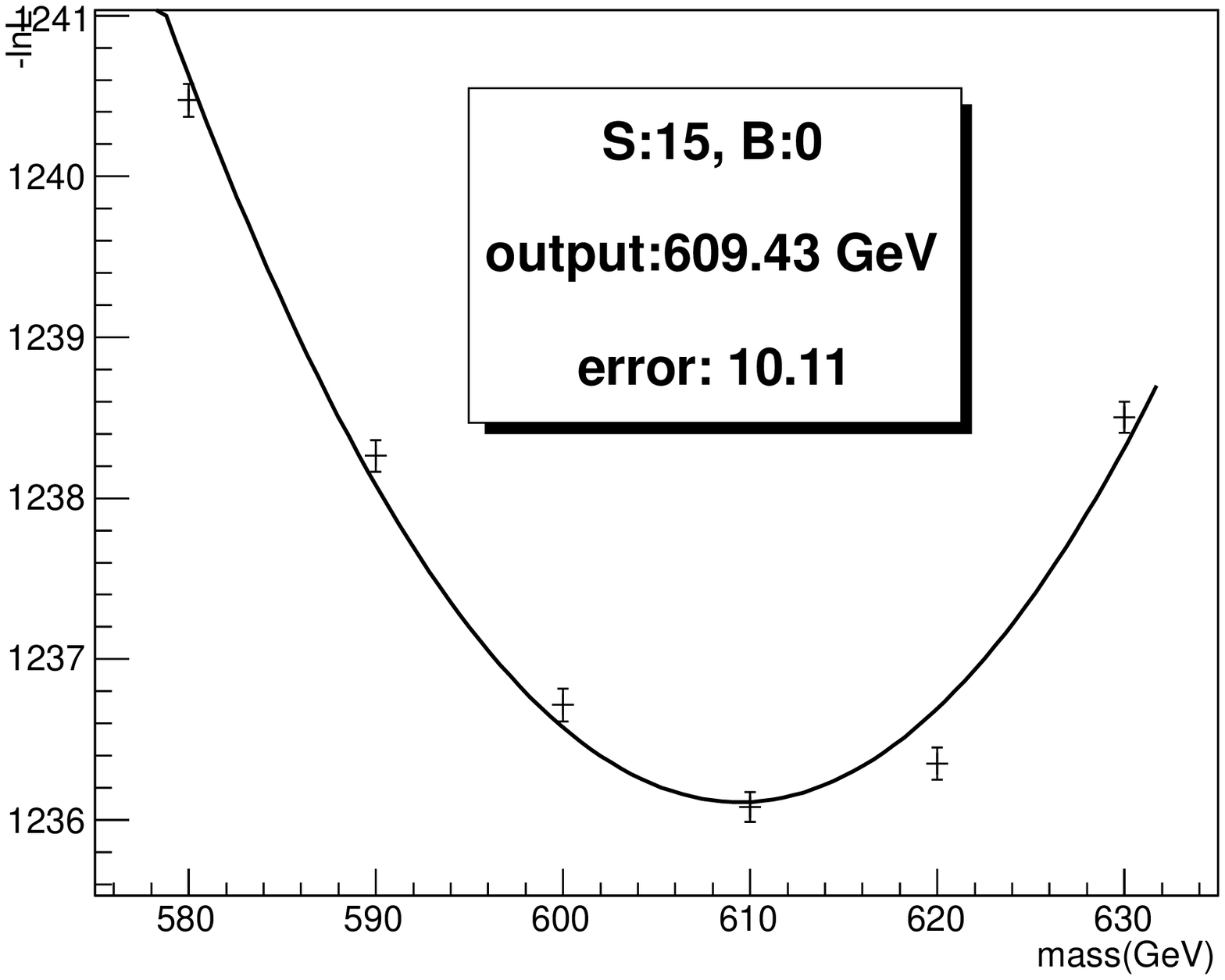}\includegraphics[scale=0.26]{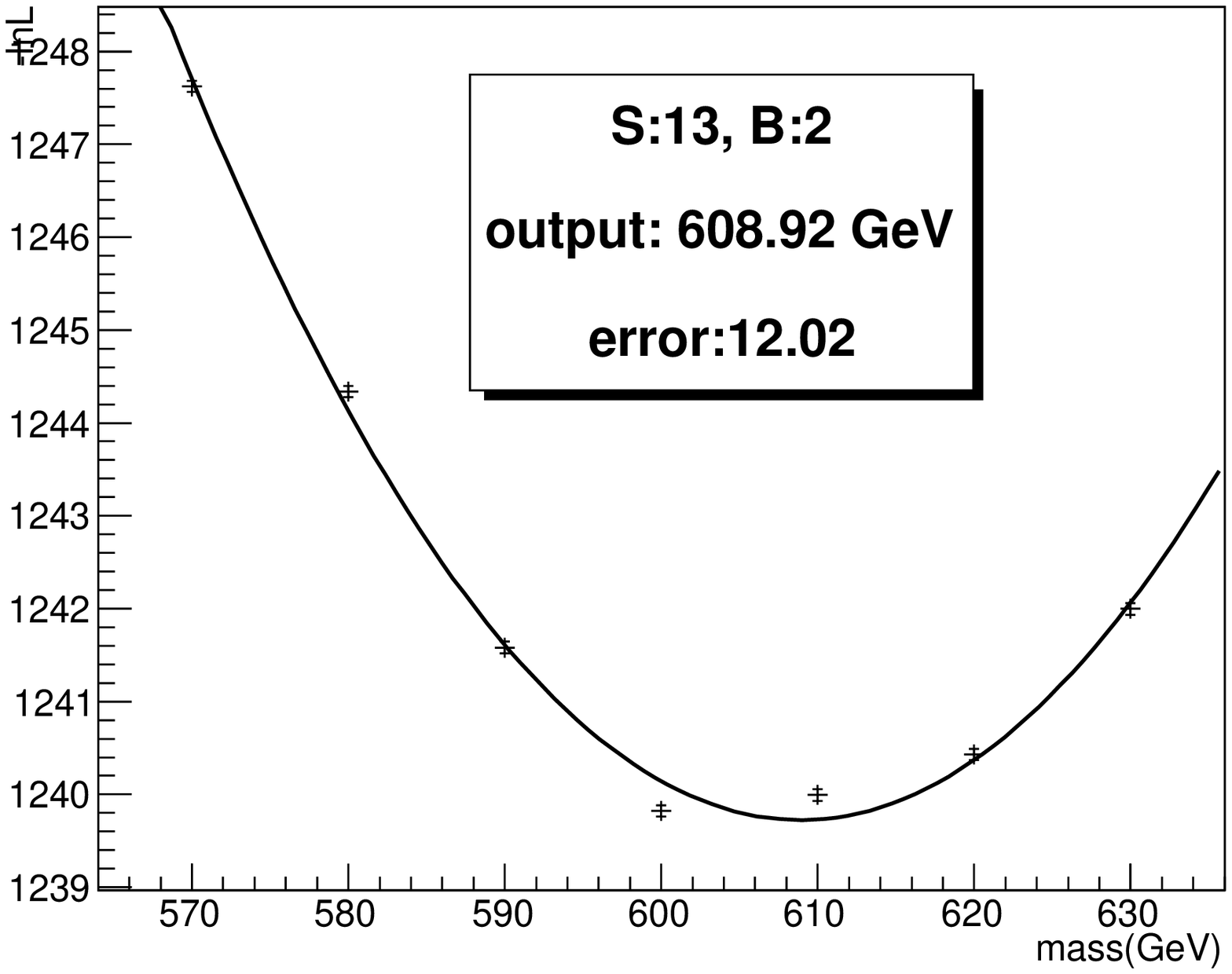}\includegraphics[scale=0.26]{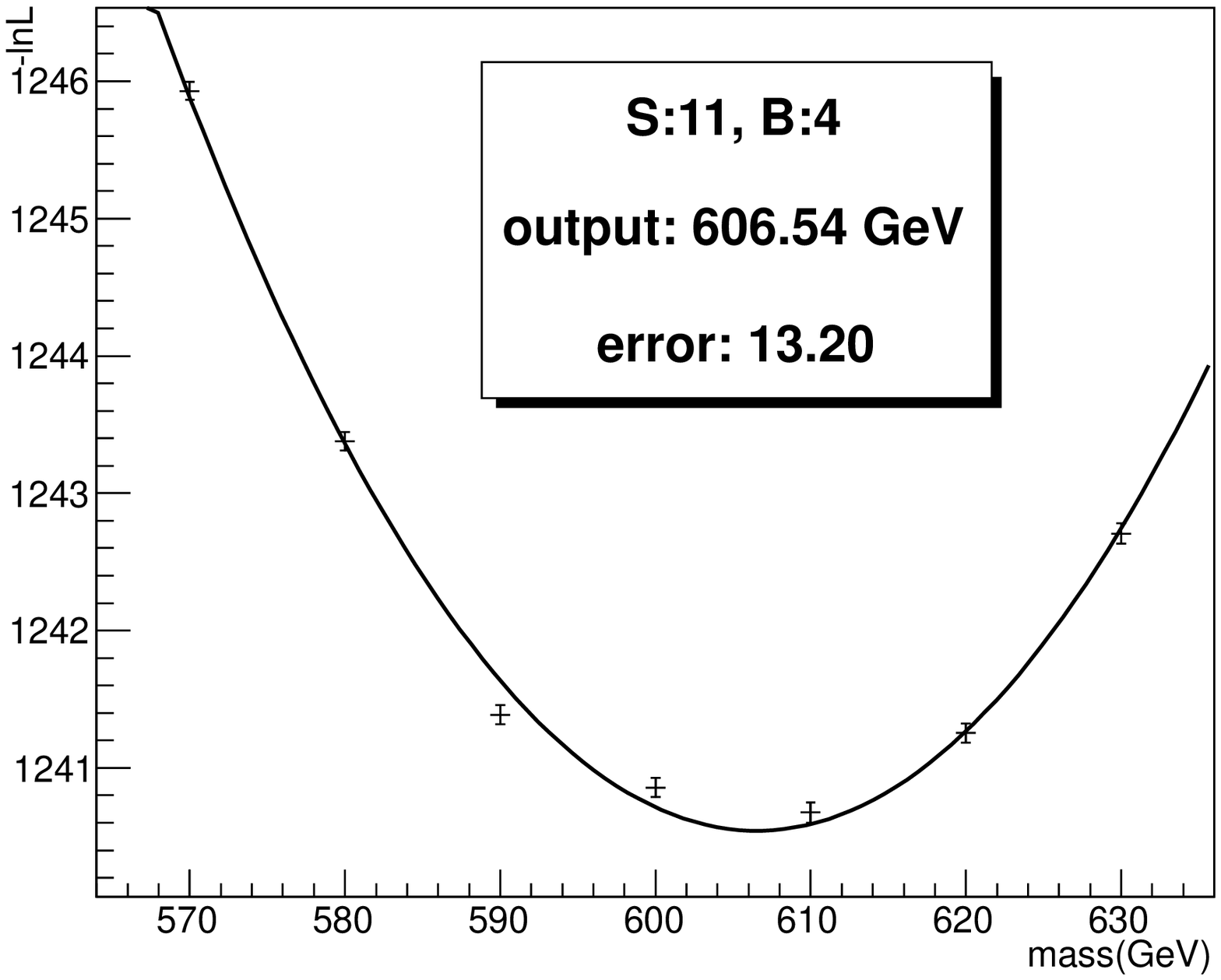}

\includegraphics[scale=0.26]{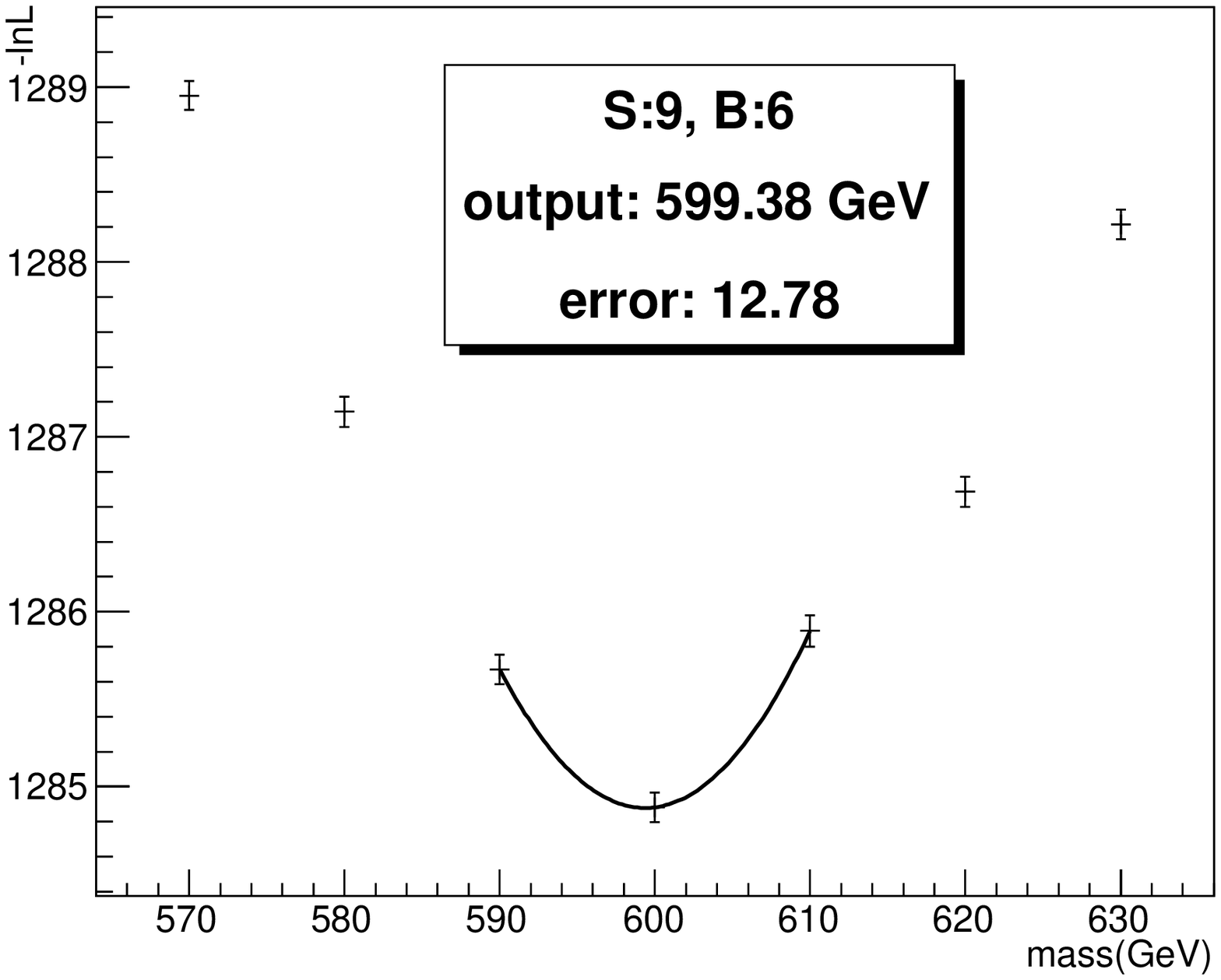}\includegraphics[width=5.2cm,height=4.4cm]{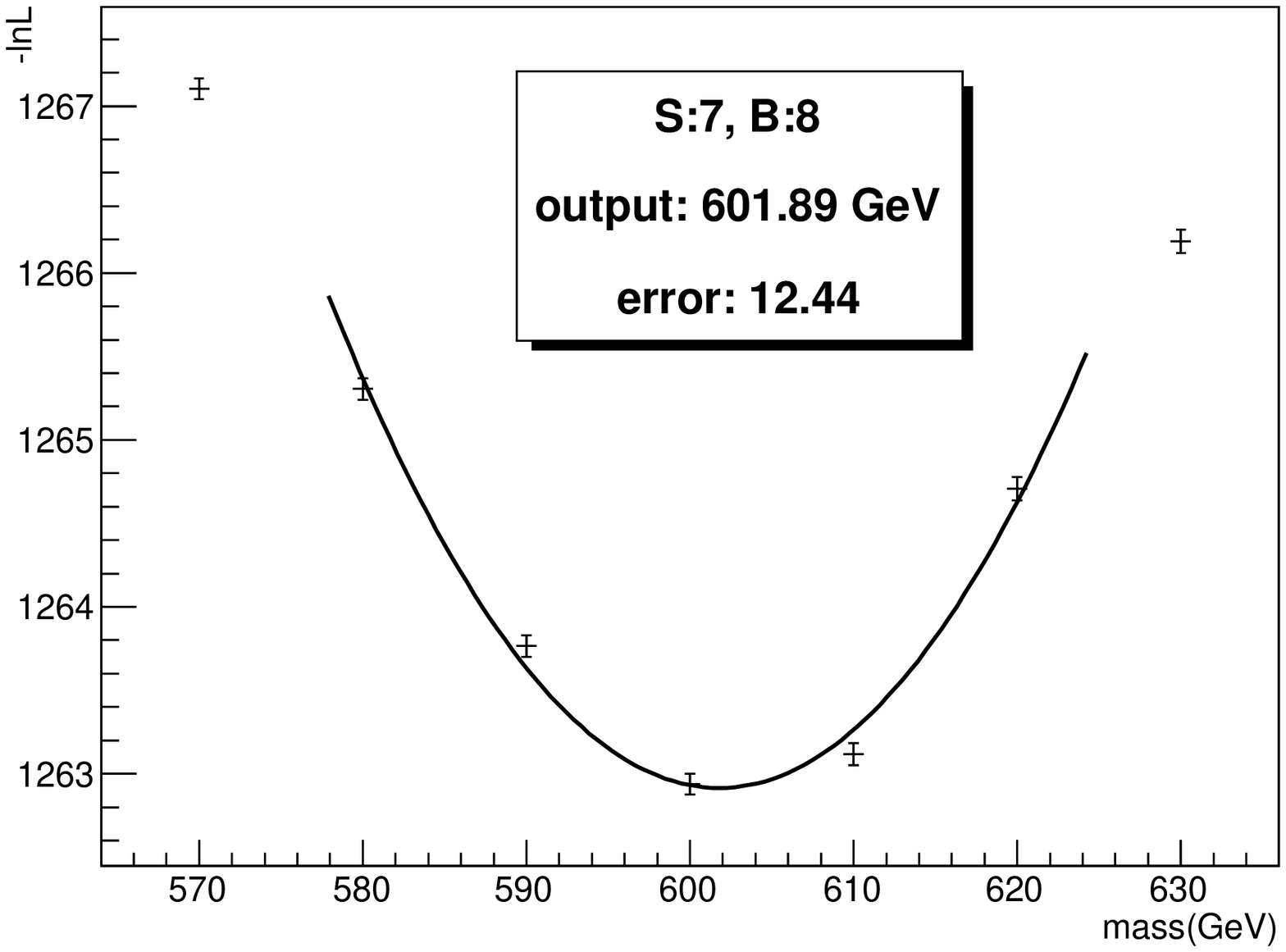}\includegraphics[width=5.2cm,height=4.4cm]{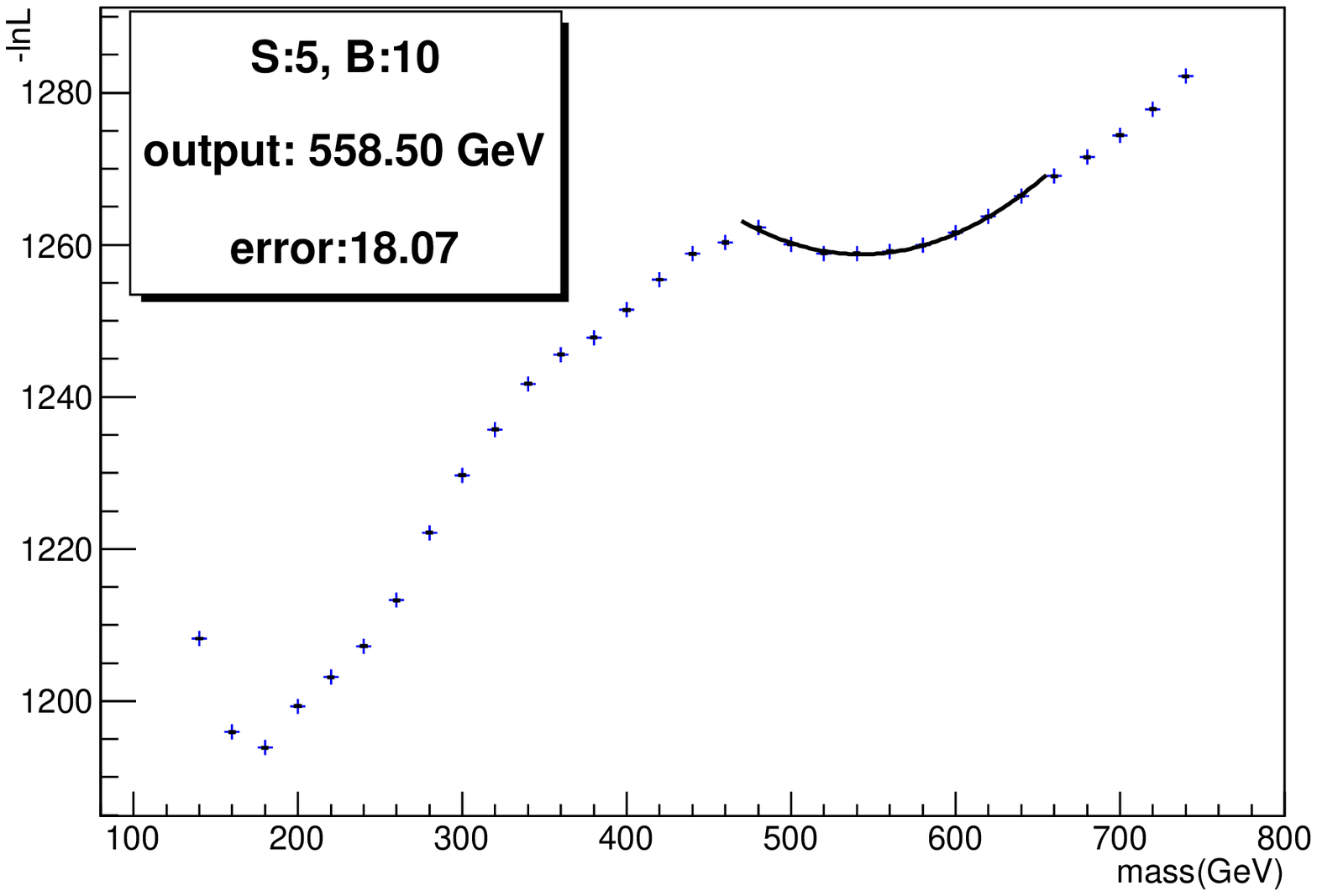}

\qquad{}\qquad{}\qquad{}\qquad{}\includegraphics[scale=0.26]{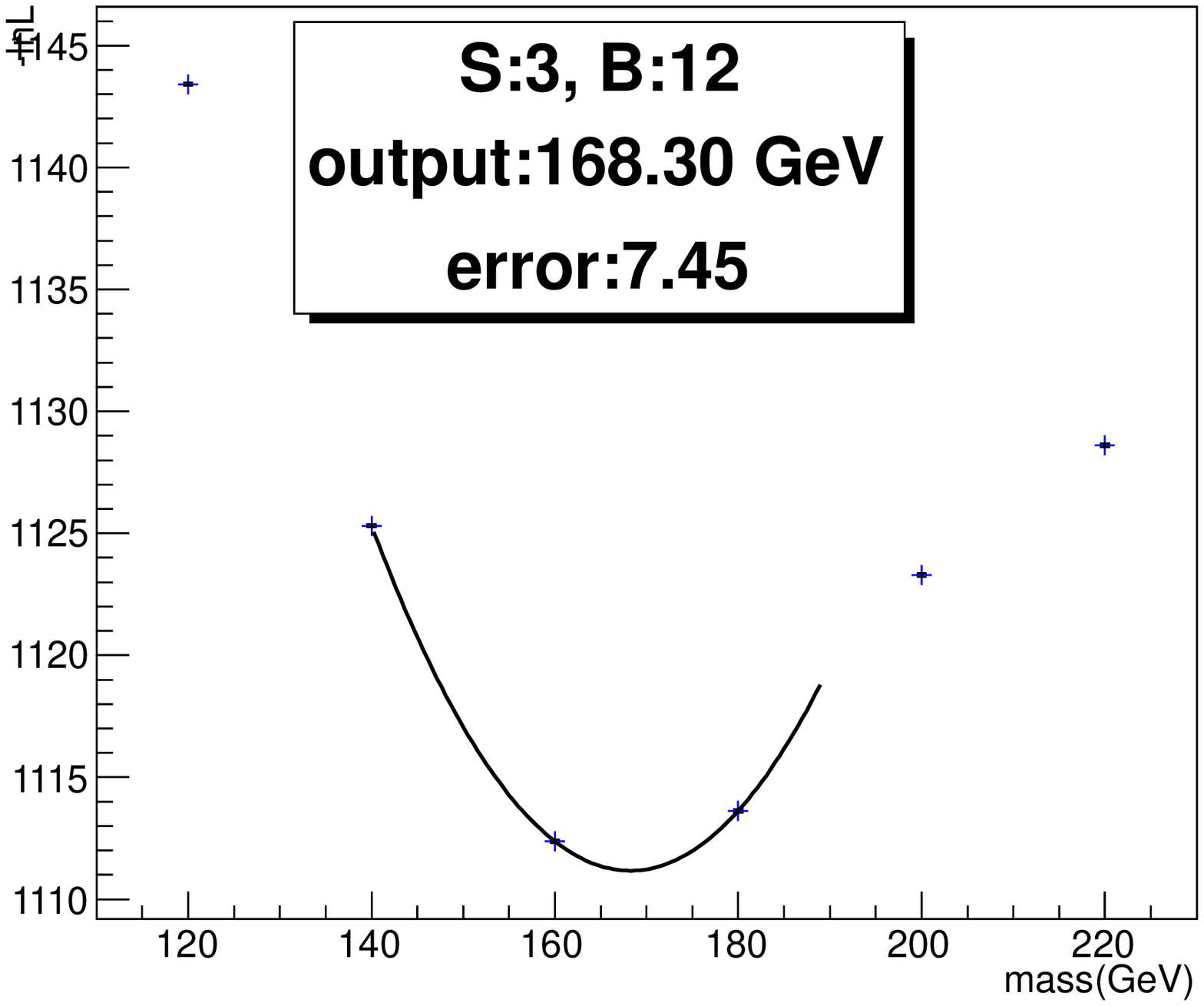}~\includegraphics[scale=0.28]{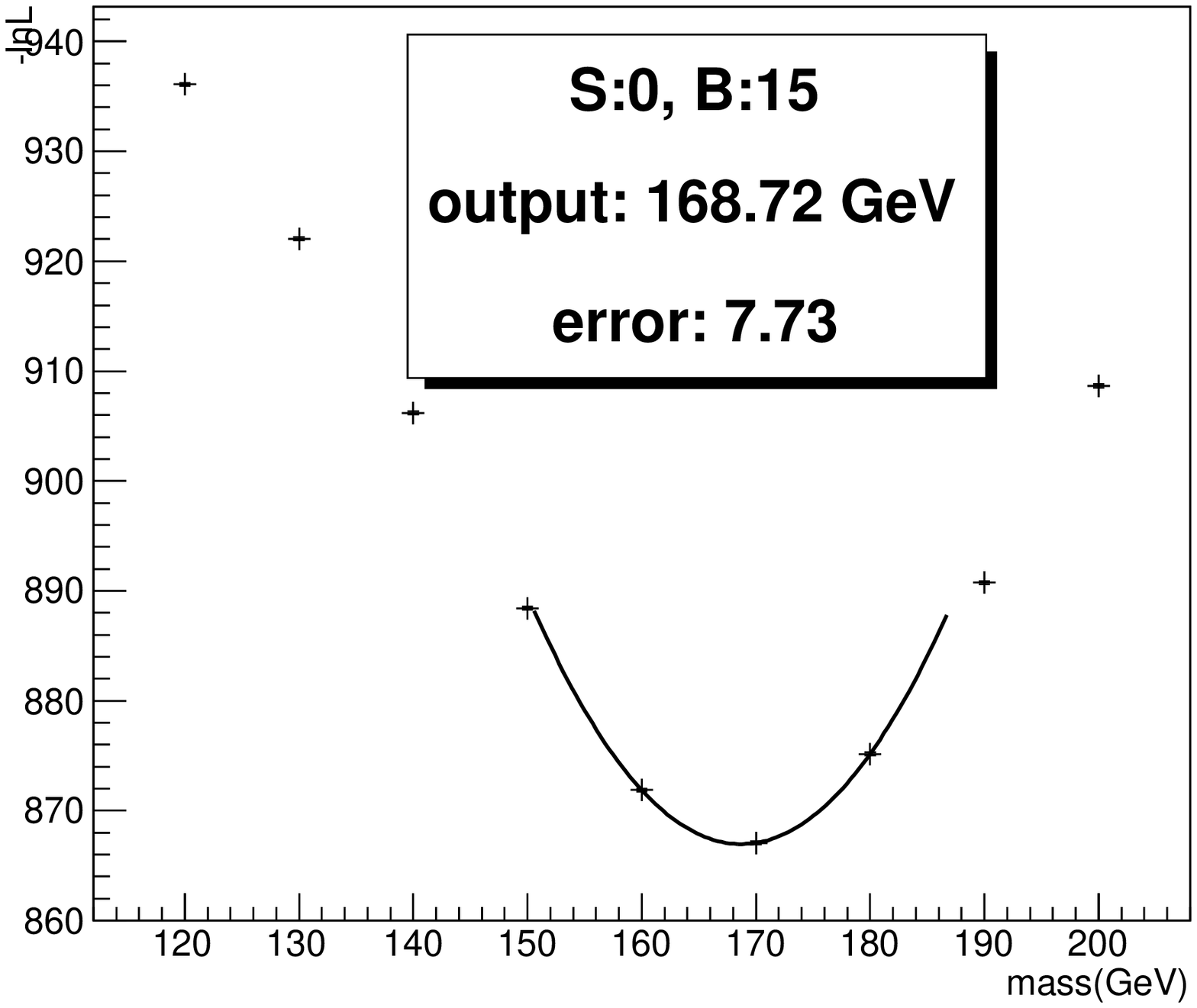}

\caption{The same as Fig. \ref{fig:500gevmem plots} but for \textit{m$_{u_{4}}$=}
600 GeV.\label{fig:600gevmem plots}}
\end{figure}

The same procedure has been applied for event samples produced with
input masses of 400 and 600 GeV. The resulting curves are shown in
Figs. \ref{fig:400gevmem plots} and \ref{fig:600gevmem plots}, respectively.
\begin{quotation}
\centering

\begin{figure}[H]
\includegraphics[scale=0.26]{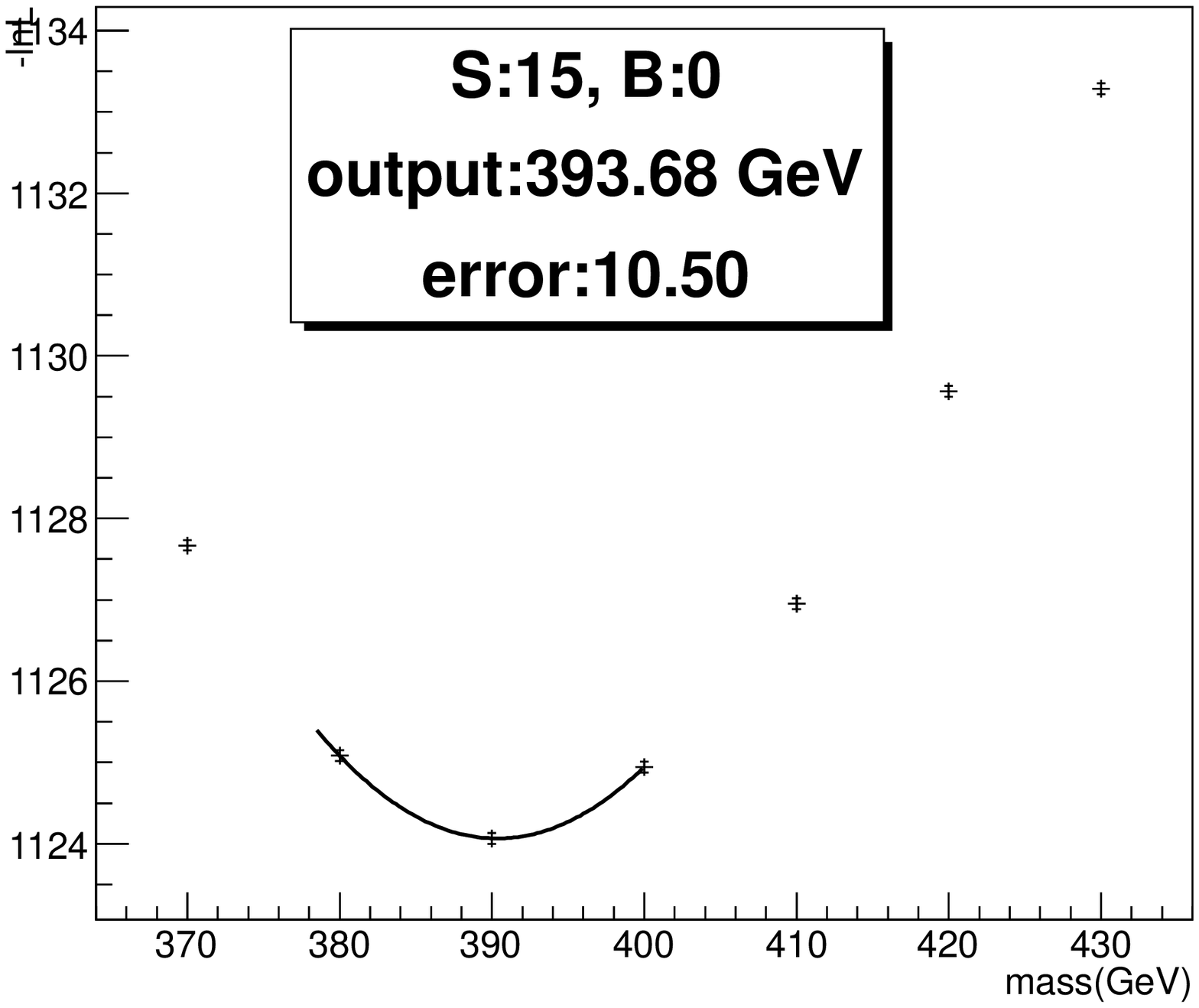}\includegraphics[scale=0.26]{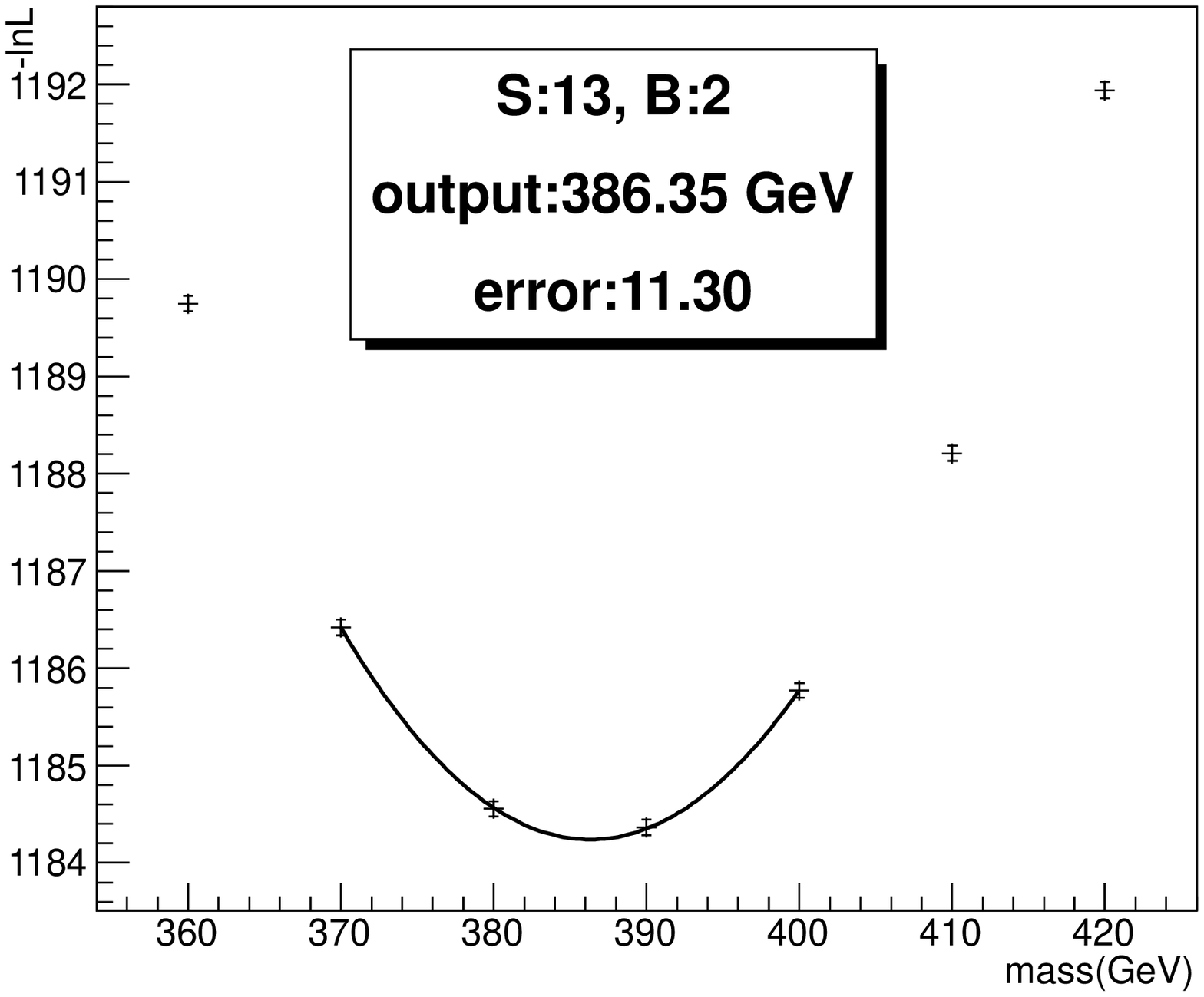}\includegraphics[scale=0.26]{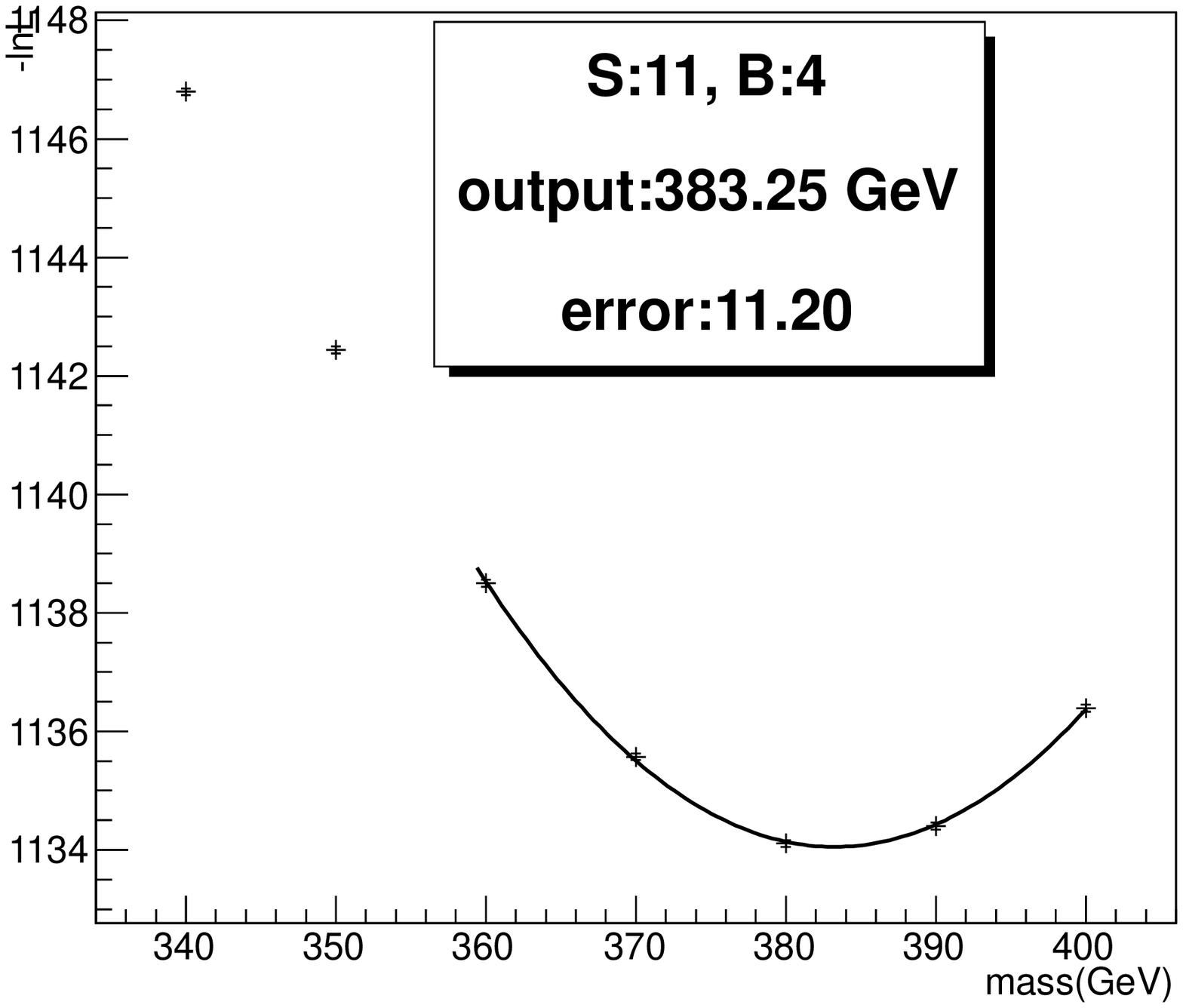}

\includegraphics[scale=0.26]{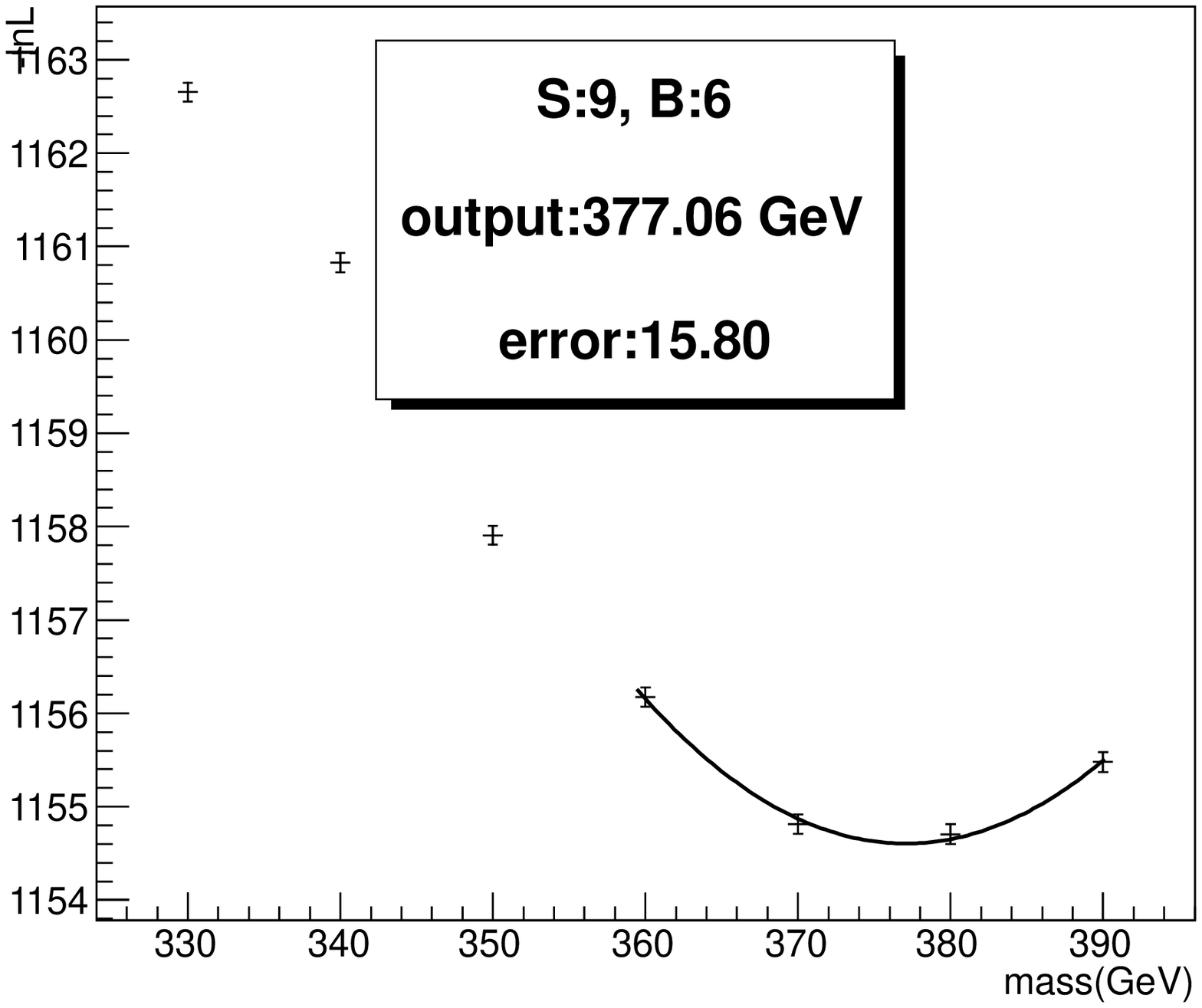}\includegraphics[scale=0.26]{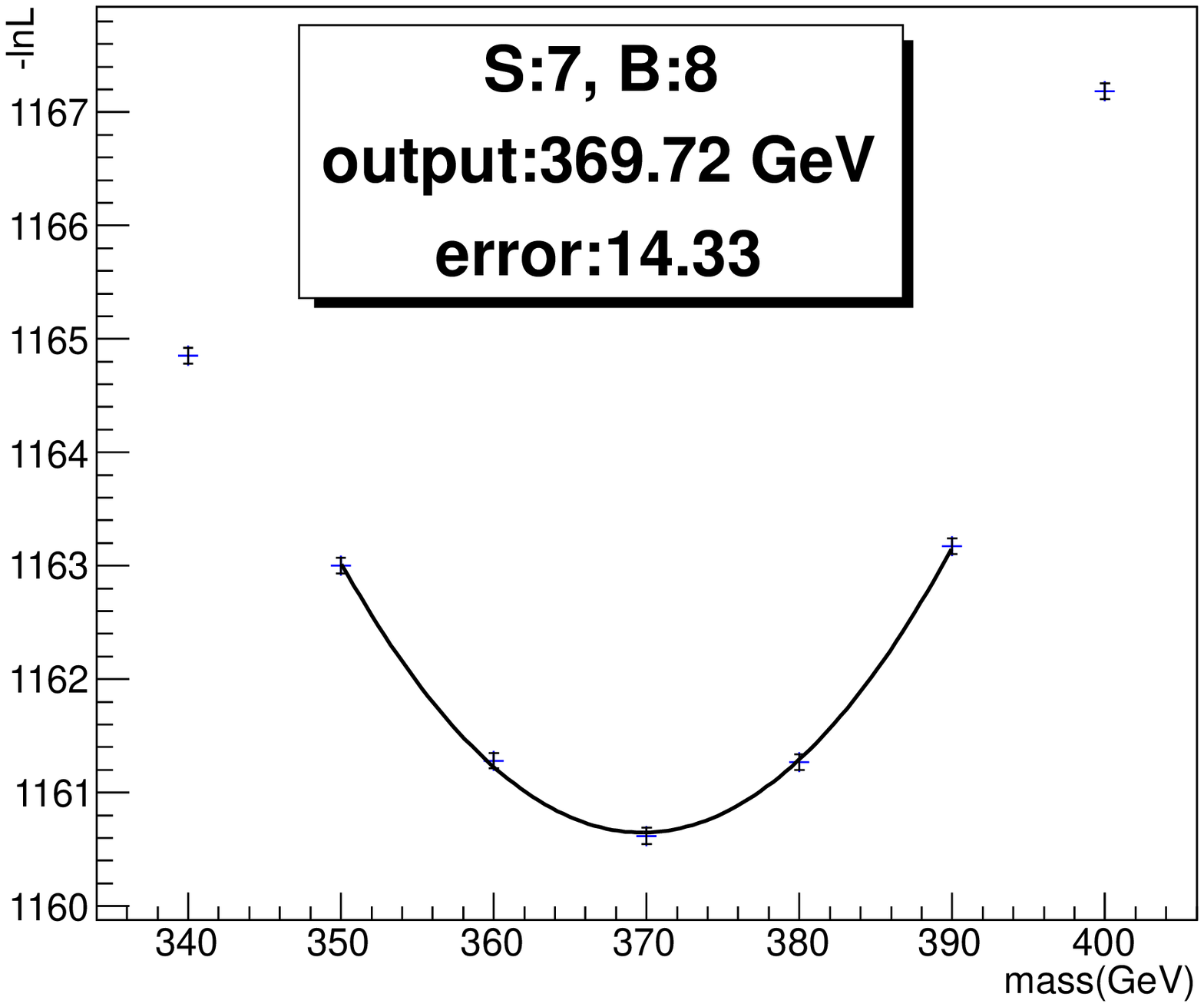}\includegraphics[width=5.2cm,height=4.4cm]{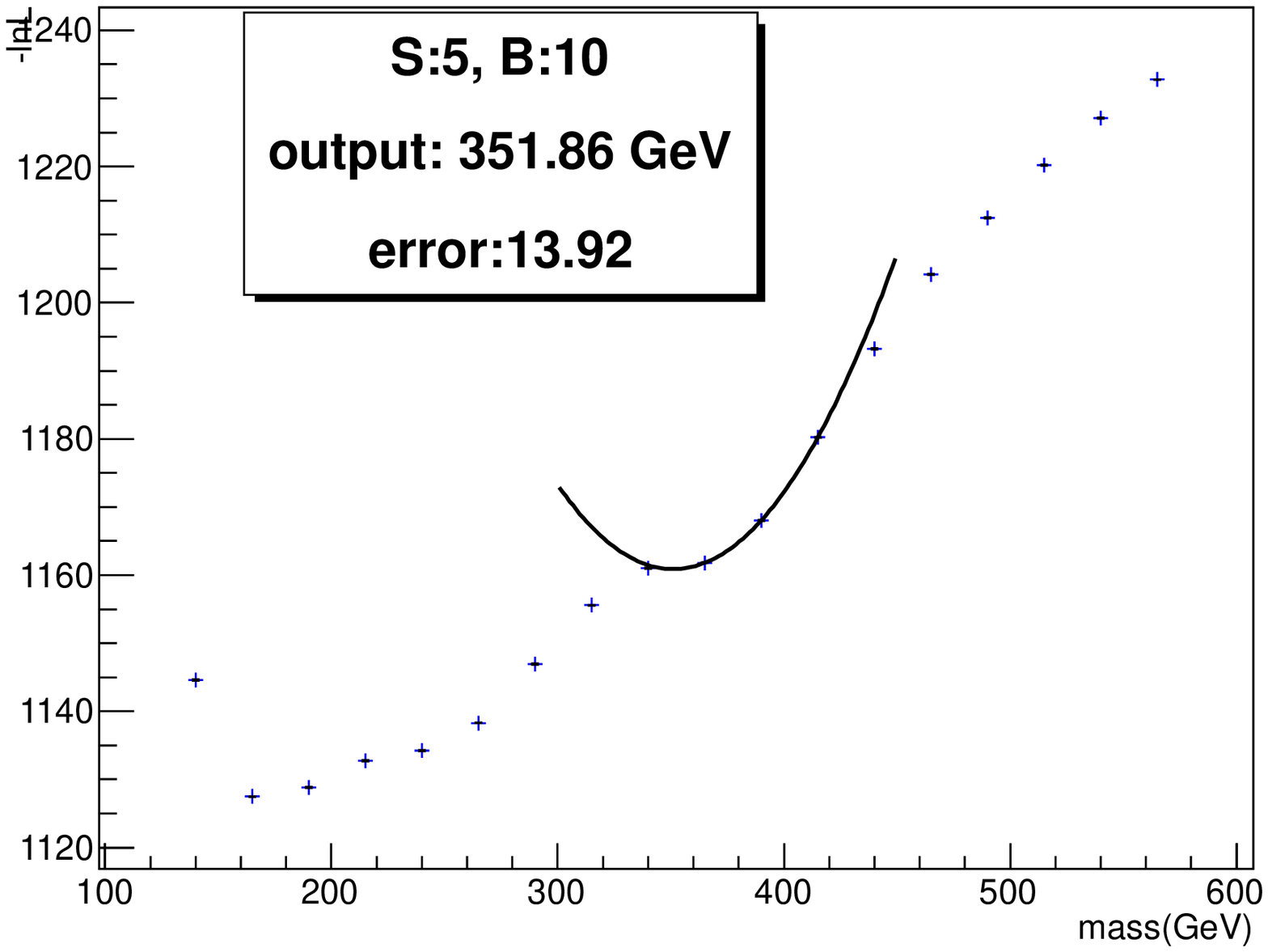}

\qquad{}\qquad{}\qquad{}\qquad{}\includegraphics[scale=0.26]{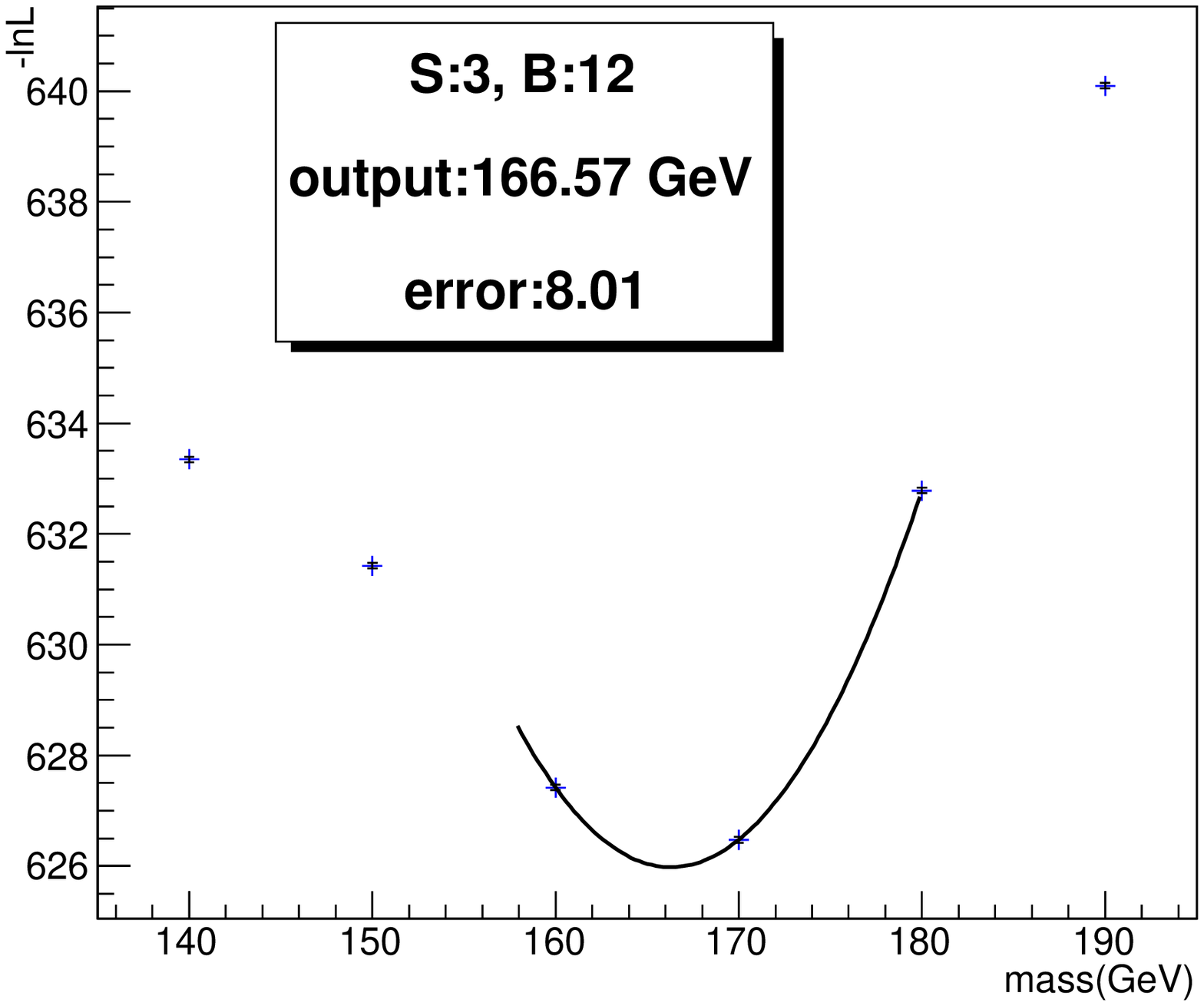}~\includegraphics[scale=0.28]{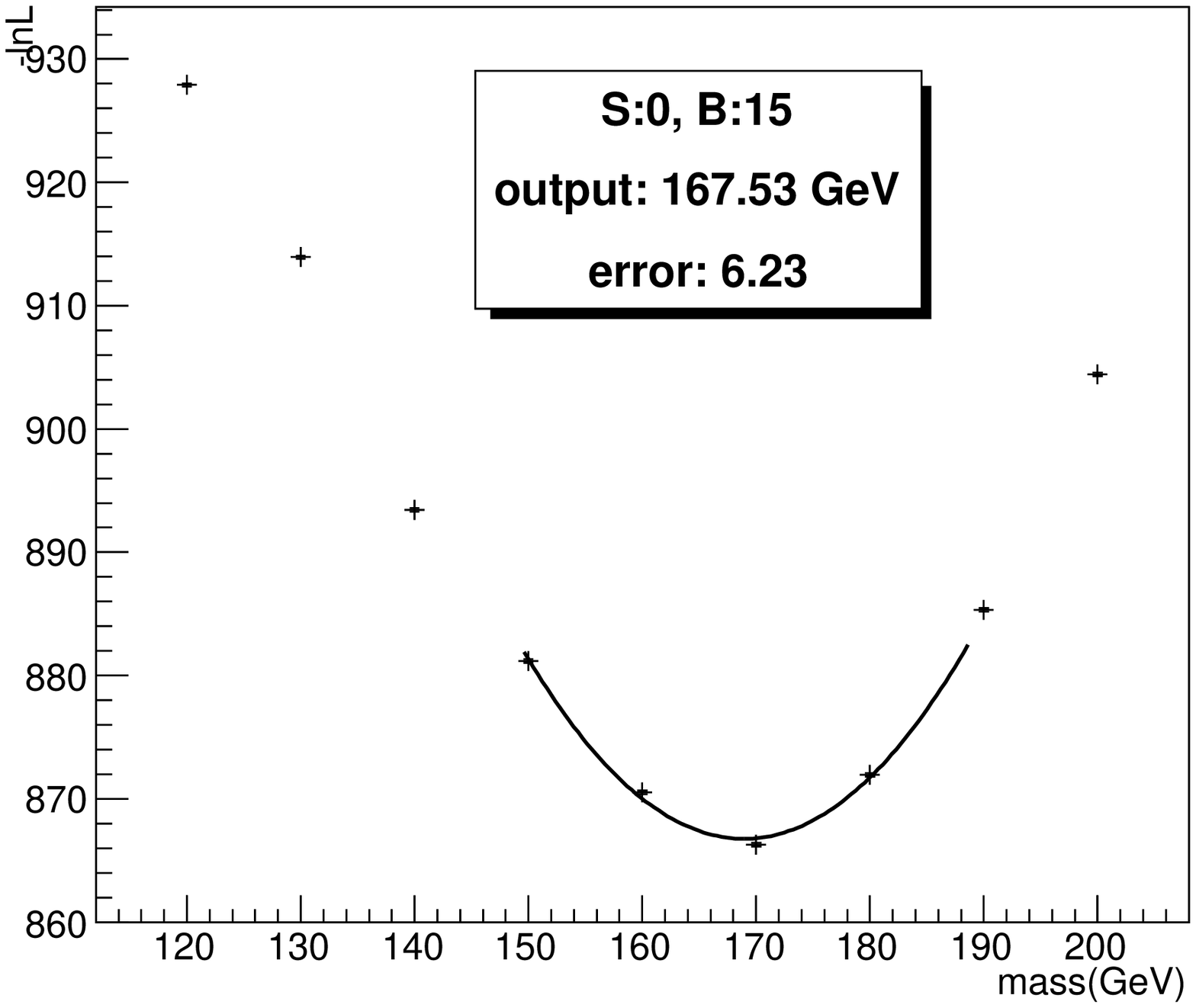}

\caption{The same as Fig. \ref{fig:500gevmem plots} but for \textit{m$_{u_{4}}$=}
400 GeV.\label{fig:400gevmem plots}}
\end{figure}

\end{quotation}

The input masses and the reconstructed masses with statistical errors
using matrix element technique from the final state with different
S/B ratios are shown in the Table \ref{tab:Matrix-Element-analysis}.

\begin{table}[H]
\caption{Matrix Element analysis results obtained for various \textit{u$_{4}$}
input masses and event samples which include various S/B ratios.\label{tab:Matrix-Element-analysis}}
\end{table}

\noindent %
\begin{tabular}{|l|c|c|c|}
\hline 
\multirow{2}{*}{Event sample} & \multicolumn{3}{c|}{Output \textit{u$_{4}$} masses for }\tabularnewline
\cline{2-4} 
 & input mass= 400 GeV & input mass= 500 GeV & input mass= 600 GeV\tabularnewline
\hline 
15 signal & \multicolumn{1}{c|}{393.68$\pm$ 10.50} & \multicolumn{1}{c|}{503.41$\pm$8.14} & 609.43$\pm$10.11\tabularnewline
\hline 
13 signal + 2 backg. & 386.35$\pm$ 11.30 & 498.91$\pm$10.04 & 608.92$\pm$12.02\tabularnewline
\hline 
11 signal + 4 backg. & 383.25 $\pm$11.20 & 499.72$\pm$11.65 & 606.54$\pm$13.20\tabularnewline
\hline 
9 signal + 6 backg. & 377.06$\pm$ 15.80 & 495.54$\pm$15.29 & 599.38$\pm$12.78\tabularnewline
\hline 
7 signal + 8 backg. & 369.72$\pm$ 14.33 & 487.43$\pm$17.31 & 601.89$\pm$12.44\tabularnewline
\hline 
5 signal + 10 backg. & 351.86$\pm$ 13.92 & 471.50$\pm$24.19 & 558.50$\pm$18.07\tabularnewline
\hline 
3 signal + 12 backg. & 166.57$\pm$ 8.01 & 167.77$\pm$8.32 & 168.30$\pm$7.45\tabularnewline
\hline 
0 signal + 15 backg. & 167.53$\pm$ 6.23 & 171.26$\pm$ 6.11  & 168.72$\pm$7.73 \tabularnewline
\hline 
\end{tabular}

~

By comparing Table \ref{tab:cbasedtablo}\&\ref{tab:Matrix-Element-analysis},
it can be clearly seen that, MEM gives much smaller deviations from
the input values for masses and errors compared to the cut-based analysis.
In addition, as number of background events increased, the resulting
value approaches the top quark mass again oppositely to the cut-based
results.

Furthermore, when the relative deviation from the true value (True
Value-Reconstructed Value/True Value) is plotted against the Signal/Signal+Background
ratio, one notices that, the deviations obtained from matrix element
method are much smaller than the ones extracted from the cut-based
analysis technique, especially for S/S+B values greater than 0.2.

\begin{figure}[H]
\centering\includegraphics[scale=0.6]{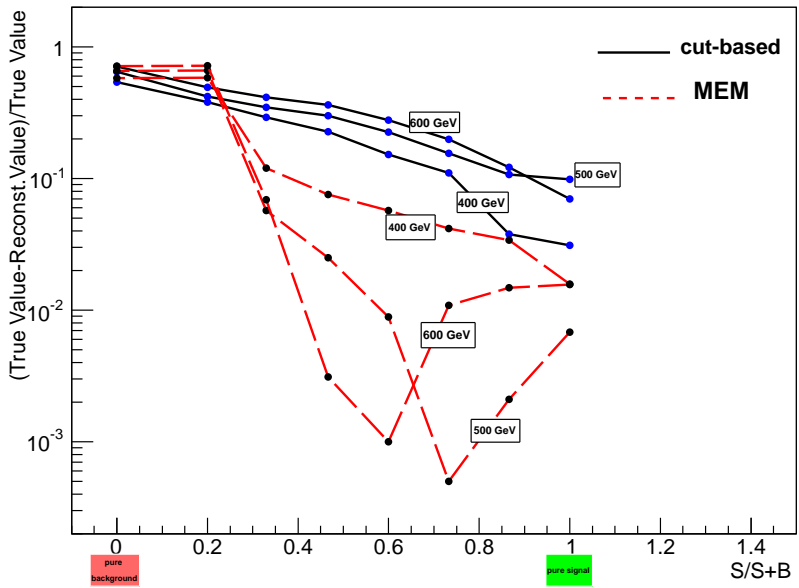}

\caption{S/S+B ratios vs corresponding relative errors for both Cut-Based and
Matrix Element Method results for different input\textit{ u$_{4}$
}masses.\label{fig:Comparison-of-S/B} }
\end{figure}

As shown in Fig. \ref{fig:Comparison-of-S/B}, matrix element method
becomes less accurate in the region of S/S+B < 0.2.

\section{Conclusion:}

This study shows that for data samples containing small number of
events with various signal to background ratios, the matrix element
method gives essentially better values for the parameter of interest
(mass of fourth family up type quark, in this analysis) and associated
statistical errors. Error values obtained from MEM are, on the average,
ten times lower than cut-based results. As a second result, MEM is,
also a powerful tool to discriminate signal and background events
even with small statistical data if S/S+B > 0.2.

~

\-

Acknowledgements: The authors would like to special thanks Saleh Sultansoy
for his very valuable comments and Olivier Mattelaer to many helpful
answers for MadWeight related issues. E.A. acknowledges the support
from the Turkish Atomic Energy Authority.

\section{References:}

\noindent {[}1{]} A. Abulencia \textit{et al.} {[}CDF Collaboration{]},
Phys. Rev. D $\boldsymbol{75}$(2007) 031105 {[}arXiv:hep-ex/0612060{]}.

\noindent {[}2{]} T. Aaltonen \textit{et al.} {[}CDF Collaboration{]},
Phys. Rev. Lett. $\boldsymbol{99}$ (2007) 182002 {[}arXiv:hep-ex/0703045{]}.

\noindent {[}3{]} V. M. Abazov \textit{et al.} {[}D0 Collaboration{]},
Phys. Rev. D $\boldsymbol{74}$ (2006) 092005 {[}arXiv:hep-ex/0609053{]}.

\noindent {[}4{]} V. M. Abazov \textit{et al}. {[}D0 Collaboration{]},
Phys. Rev. D $\boldsymbol{75}$ (2007) 092001 {[}arXiv:hep-ex/0702018{]}.

\noindent {[}5{]}V. M. Abazov et al., Phys. Rev. Lett. 103 (2009)
092001.

\noindent {[}6{]} V. M. Abazov et al., \textquotedblleft{}Helicity
of the W boson in lepton + jets tt\={ }events\textquotedblright{},
Phys. Lett., vol. B617, pp. 1\textendash{}10, 2005, hep-ex/0404040.

\noindent {[}7{]} T. Aaltonen et al., Phys. Rev. Lett. 103 (2009)
092002.

\noindent {[}8{]} K. Kondo, J. Phys. Soc.Jap. $\boldsymbol{57}$ (1988)
4126.

\noindent {[}9{]} R. H. Dalitz and G. R. Goldstein, Phys. Rev. D $\boldsymbol{45}$
(1992) 1531.

\noindent {[}10{]} P. Artoisenet, O. Mattelaer, MadWeight: automatic
event reweighting with matrix elements, Prospects for Charged Higgs
Discovery at Colliders, September 16-19 2008, Upsala, Sweden.

\noindent {[}11{]} MadWeight, \textit{\uline{\href{https://server06.fynu.ucl.ac.be/projects/madgraph/wiki/MatrixElement}{https://server06.fynu.ucl.ac.be/projects/madgraph/wiki/MatrixElement}}}

\noindent {[}12{]} J. C. E. Vigil, Maximal Use of Kinematic Information
for the Extraction of the Mass of the Top Quark in Single-lepton tt
events at DØ, Univ. of Rochester, New York, 2001.

\noindent {[}13{]} M. F. Canelli. Helicity of the W boson in single-lepton
tt\={ } events, FERMILAB-THESIS-2003-22 (2003).

\noindent {[}14{]} O. Mattelaer, A New Approach to Matrix Element
Re-Weighting, Universite Catholique de Louvain, Centre for Cosmology,
Particle Physics and Phenomenology, January 2011.

\noindent {[}15{]} B. Holdom, W.S. Hou, T. Hurth, M. Mangano, S. Sultansoy,
G. Ünel, Four Statements about the Fourth Generation, PMC Phys. A3
(2009) 4. 

\noindent {[}16{]} F. Maltoni and T. Stelzer, JHEP 0302 (2003) 027
{[}arXiv:hep-ph/0208156{]}.

\noindent {[}17{]} T. Sjostrand, S. Mrenna and P. Skands, JHEP 0605
(2006) 026 {[}arXiv:hep-ph/0603175{]}.

\noindent {[}18{]} J. Conway, \textit{\uline{\href{http://physics.ucdavis.edu/~conway/research/software/pgs/pgs4-general.htm}{http://physics.ucdavis.edu/$\sim$conway/research/software/pgs/pgs4-general.htm}}}

\noindent {[}19{]} J. Pumplin \textit{et al.}, \textquotedblleft{}New
generation of parton distributions with uncertainties from global
QCD analysis\textquotedblright{}, JHEP, vol. 07, p. 012, 2002, {[}arXiv:hep-ex/0201195{]}.

\noindent {[}20{]} LHCO wiki page, \href{http://www.jthaler.net/olympicswiki/doku.php}{http://www.jthaler.net/olympicswiki/doku.php}
\end{document}